%% file: mainABC.tex
\documentclass[aps,prb,twocolumn,superscriptaddress,showpacs,longbibliography]{revtex4-2}
\input{packages}

\begin{document}

\title{
High-Efficiency Thermoelectric Transport in Aharonov-Bohm-Casher Rings
}
\author{Diego García\hspace{0.08cm}\orcidlink{0009-0006-9595-5468}}
\affiliation{Institut de F\'{\i}sica Interdisciplin\`aria i Sistemes Complexos
IFISC (CSIC-UIB), E-07122 Palma de Mallorca, Spain}
\affiliation{Physics Department, Lund University, SE-22100 Lund, Sweden}

\author{Sergio Arias\hspace{0.08cm}\orcidlink{0009-0004-4458-7402}}
\affiliation{Institut de F\'{\i}sica Interdisciplin\`aria i Sistemes Complexos
IFISC (CSIC-UIB), E-07122 Palma de Mallorca, Spain}
\affiliation{Department of Mathematics, Universidad Carlos III, Madrid, Spain}

\author{Rosa López\hspace{0.08cm}\orcidlink{0000-0002-3717-6347}}
\affiliation{Institut de F\'{\i}sica Interdisciplin\`aria i Sistemes Complexos
IFISC (CSIC-UIB), E-07122 Palma de Mallorca, Spain}

\begin{abstract}
Quantum heat engines are nanoscale devices that convert heat into work by exploiting quantum effects, such as coherence and interference. Previous studies of these devices did not consider spin-dependent effects, which can influence the thermoelectric performance of the engine. In this work, we study the thermoelectric behavior of a quantum heat engine based on an Aharonov-Bohm ring -- a mesoscopic ring where electrons exhibit interference depending on the magnetic flux it encloses -- incorporating Rashba spin-orbit interaction (SOI), which couples the electron's motion and spin. We find that Rashba SOI enhances the figure of merit $ZT$, measure of the engine's conversion efficiency. Our results suggest that controlling spin-dependent interference could lead to improvements in the fabrication of efficient thermoelectric devices. 
\end{abstract}

\maketitle
\input{Sections/SectionIntroductionRosa}
\input{Sections/SectionModelRosa}
\input{Sections/SectionScatteringRosa}

\input{Sections/SectionTransportRosa}
\input{Sections/SectionResultsRosa}
\input{Sections/SectionVIConclusionsRosa}

\section*{Acknowledgements}
R.L. acknowledges support by the Spanish State Research Agency (No. MCIN/AEI/10.13039/501100011033) and FEDER (UE) under Grants No. PID2020-117347GB-I00 and No. PID2023-151975NB-I00 and María de Maeztu Project No. CEX2021-001164-M.

\twocolumngrid
\bibliographystyle{apsrev4-1}
\bibliography{bibliography}

\end{document}

%% file: packages.tex
\usepackage{graphicx}
\usepackage{amsmath,amssymb}
\usepackage{physics}
\usepackage{float}
\usepackage[utf8]{inputenc}
\usepackage{natbib}
\usepackage{etoolbox}
\usepackage{nicefrac}
\usepackage[percent]{overpic}
\usepackage[colorlinks=true, linkcolor=magenta, citecolor=magenta, urlcolor=magenta]{hyperref}

\usepackage{subfigure,overpic}
\usepackage{comment}
\usepackage{comment}
\usepackage{dsfont}
\usepackage{booktabs}

\usepackage{hyperref}
\usepackage{academicons} 
\usepackage{xcolor}      

\newcommand{\orcid}[1]{\href{#1}{\textcolor{lime!60!black}{\aiOrcid}}}
\usepackage{orcidlink}

%% file: Sections/SectionIntroductionRosa.tex
\section{Introduction}

One of the major breakthroughs in thermodynamics was achieved by N. Sadi Carnot, who, in 1824, established the foundations for heat engines and energy conversion. More recently, these ideas have evolved into the field of quantum thermodynamics, which has attracted significant attention from scientists across a wide range of fields due to the many applications it offers \cite{Giazz,Benenti,Pekola,Potanina,Sothmann}. The study of quantum systems as thermal machines exploits features such as coherence and entanglement, which are completely absent in classical devices, thereby offering opportunities to surpass conventional performance bounds \cite{Blasi,Ralph,Hwarn,Mojtaba}.

Thermoelectric transport -- the coupling between heat and charge currents -- is a central topic in quantum device research. It enables both the generation of electrical power from temperature gradients and the manipulation of thermal flows using electronic control. In mesoscopic  conductors, where the system size is comparable to the electron phase-coherence length ($l_\phi$) \cite{Beenakker,Datta}, classical thermodynamics still holds, but quantum interference and phase coherence have a bigger impact on thermoelectric behavior. Such systems exhibit new rich phenomena: their transmission properties and energy dependence can enhance thermopower and reveal new interference patterns and effects. Therefore, studying thermoelectricity in mesoscopic systems could have potential applications in the design of advanced heat engines, refrigerators, and novel energy-harvesting devices.

A central focus of modern thermoelectric research is the control of geometric and electronic asymmetry in device architectures. Asymmetric nanostructures -- such as quantum rings with unequal arm lengths -- can create strong energy-dependent modulations in the transmission function, affecting both charge and heat transport. By controlling the placement of the scatterers or the length of the ring's arms, it is possible to access new interference patterns and spin-resolved transport effects.
\begin{figure}[h!]

  \centering
  \includegraphics[width=0.5\textwidth]{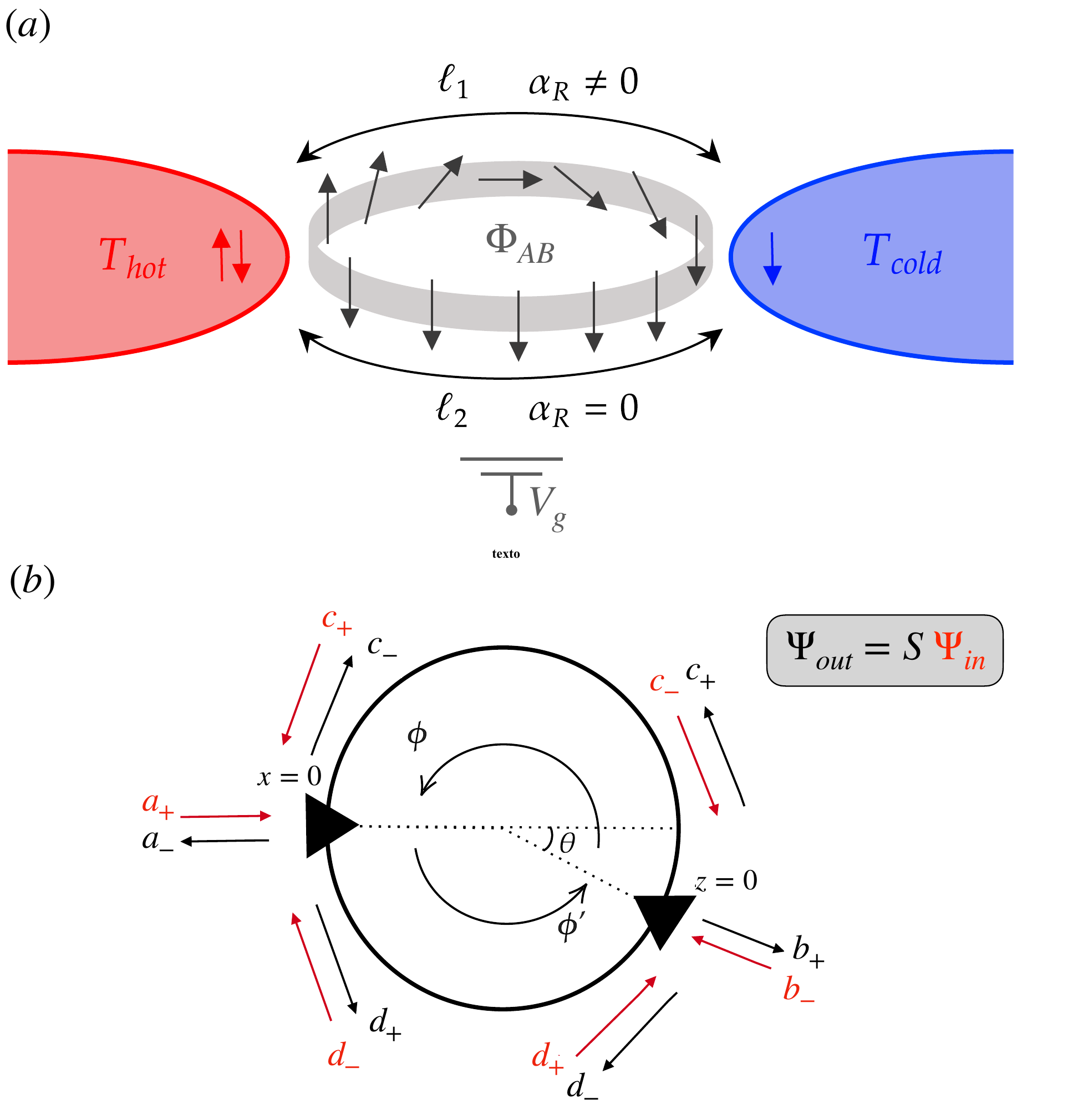}

  \caption{(a) Schematic representation of a two-terminal Aharonov-Bohm quantum heat engine. The upper arm, of length $\ell_1$, is affected by Rashba SOI, while the lower arm, of length $\ell_2$, remains spin-independent, as indicated. The ring encloses a magnetic flux $\Phi_{AB}$ and is connected to two reservoirs at temperatures $T_{\rm hot}$ and $T_{\rm cold}$ through junctions of coupling strength $\epsilon$. $V_g$ represents the applied gate voltage. (b) Parametrization of the AB ring, where the junctions act as scatterers. At each junction, the incoming (red) and outgoing (black) wavefunctions are related by the scattering matrix $\mathcal{S}$.}
  \label{fig:ABinter}
  \vspace{-0.3cm}
\end{figure}

The \textit{Aharonov-Bohm (AB) effect} \cite{AB}, in which a charged particle accumulates a magnetic phase when traveling around an area enclosing a magnetic flux, provides a theoretical and experimental framework for analyzing coherent transport. AB rings enable the study of phase-dependent transmission and thermoelectric response and have been proposed as promising platforms for \textit{quantum heat engines} \cite{Gérald,Behera,Bedkiha}. The addition of spin-dependent effects -- specifically, \textit{Rashba spin-orbit interaction (SOI)} \cite{Rashba}, which couples the electron's spin to its momentum -- further enriches this scenario. This effect can be tuned electrostatically via the gate voltage and leads to spin precession and spectral splitting. When Rashba SOI is only present in one arm and scatterers are deliberately placed at different locations, the quantum ring becomes a versatile platform for exploring the interplay of spatial, spin, and quantum coherence effects on thermoelectric transport.

This work investigates in detail the thermoelectric response of an asymmetrically configured two-terminal AB quantum ring heat engine, where Rashba SOI is present in only one arm and the placement of the scatterers breaks rotational symmetry (see Fig.~\hyperref[fig:ABinter]{1}). In this configuration, electrons can travel through two coherent paths and acquire different spin-dependent phases, resulting in interference effects that influence transport. The ability to modulate these properties by varying spatial configuration and external controls is central to optimizing quantum thermal machines, both for efficiency and functionality. By tuning different system parameters --  particularly, the Rashba strength -- one can enhance the device's thermoelectric response, achieving higher values of the figure of merit $ZT$, a dimensionless measure of the system's energy conversion efficiency. This highlights Rashba SOI as a promising platform for designing efficient and tunable quantum heat engines.

The remainder of the paper is structured as follows: Section \ref{sec:model} introduces the theoretical model describing the AB ring with Rashba SOI and defines the relevant eigenvalues and eigenstates; Section~\ref{scattering} formulates the quantum scattering problem using the Landauer-Büttiker approach, and explains the procedure used to compute the spin-resolved transmission probabilities. Section~\ref{transport} introduces thermoelectric transport theory in the \textit{linear transport regime}, defining the main thermoelectric coefficients. Finally, Section~\ref{sec:results} presents and analyzes the results, highlighting the effect of Rashba SOI on thermoelectric behavior and the device's efficiency.

%% file: Sections/SectionModelRosa.tex
\section{Model} \label{sec:model}
We examine a one-dimensional Aharonov--Bohm quantum ring, where the electron's motion is confined to the $XY$ plane and restricted to a circular path of radius $b$. A perpendicular magnetic field $\vec{B} = B \hat{z}$ is applied along the $z$-axis, resulting in a magnetic flux $\Phi$ threading the ring. Under these conditions, the electron's motion can be described by the azimuthal angle $\phi$, with eigenfunctions $\psi_n(\phi) = e^{in\phi}$, where $n$ is an integer. The corresponding energy levels are $E_n = \Omega \left(n + \Phi/\Phi_0\right)^2$, with $\Omega=\hbar^2/2mb^2$ and $\Phi_0 = 2\pi\hbar/e$ the Dirac magnetic flux quantum. States with $n < 0$ and $n > 0$ represent, respectively, clockwise and counterclockwise orbital motion, and the symmetry relation $E_n(\Phi) = E_{-n}(-\Phi)$ highlights that reversing the magnetic flux ($\Phi \rightarrow -\Phi$) is equivalent to reversing the electron's angular momentum ($n \rightarrow -n$).

We now include the Rashba SOI contribution, which is given by \cite{Meijer}
\begin{equation} 
H_{SO} = \alpha_R \vec{\sigma} \cdot [\vec{E}\times (\vec{p}-e\vec{A})], \label{eq:3}
\end{equation}
where $\alpha_R$ is the spin-orbit coupling strength (units: eV $\cdot$ m), tunable via an external electric field assumed to point along the $z$ direction. $\vec{\sigma}$ denotes the Pauli matrices. This term couples the electron's spin to its orbital motion and gives rise to spin-dependent phases in the ring. Following Refs. \cite{Meijer} and \cite{Molnár}, the total Hamiltonian, also known as the Aharonov-Bohm-Casher Hamiltonian, simplifies to
\begin{equation}
H = \Omega\left(- i \frac{\partial}{\partial \phi} +\frac{\Phi}{\Phi_0} + \frac{\hbar \omega_{so}}{2 \Omega}\sigma_r\right)^2 \equiv \Omega\hat{\mathcal{O}}^2,
\label{eq:4}
\end{equation}
with $\hat{\sigma}_r = \cos(\phi)\sigma_x+\sin(\phi)\sigma_y$, and $\omega_{so}=\alpha_R/\hbar b$.
In Eq. (\ref{eq:4}) we forsook a constant $\hbar^2 \omega_{so}^2/4\Omega$, which does not affect the dynamics of the system. The eigenstates of the Hamiltonian in Eq.~\eqref{eq:4} are obtained by solving $\hat{\mathcal{O}}\psi(\phi) = \lambda \psi(\phi)$ with the ansatz $\psi(\phi) = e^{in \phi} \chi^T (\phi) = e^{in\phi}\left( a, b e^{i \phi}\right)$. By doing this, the energy spectrum becomes
\begin{equation}
    E_n = \Omega\lambda^2 = \Omega\left(n + \frac{\Phi}{\Phi_0}-\frac{\Phi_{AC}^{(i)}}{2\pi}\right)^2, \:\:i =\{1,2\} ,\label{eq:2.7}
\end{equation}
where $i=1,2$ labels the two spin-split branches arising from the Rashba SOI, and $\Phi_{AC}^{(i)}$ is the Aharonov-Casher phase~\cite{AB-Cash} associated to each branch. The corresponding normalized eigenstates are
\begin{equation}
\chi^{(1)} = \mathcal{N}\begin{pmatrix}
\gamma \\
\eta e^{i\phi}
\end{pmatrix}, \quad\chi^{(2)} =  \mathcal{N} \begin{pmatrix}
\eta \\
-\gamma e^{i\phi}
\end{pmatrix}, \label{eq:2.8}
\end{equation}
\noindent
where $\gamma = \hbar \omega_{so}/(2\Omega)$, and $\mathcal{N} = 1/\sqrt{\gamma^2 + \eta^2}$ is the normalization factor. The parameter $\eta = 1/2 - \sqrt{1/4 + \gamma^2} = -\Phi_{AC}^{(1)}/(2\pi)$, is then related to the Aharonov-Casher phase such that $\eta = -\Phi_{AC}^{(1)}/(2\pi)$. Upon introducing the Rashba field, the symmetry $E_n(\Phi) = E_{-n}(-\Phi)$ no longer holds, and clockwise and counterclockwise propagating states acquire different energies.

The presence of Rashba SOI mixes the spin components along the $x$ and $y$ axes, resulting in \textit{spin precession}. The eigenstates are coherent superpositions of $\chi_\uparrow^T = (1,0)$ and $\chi_\downarrow^T = (0,1)$ (with $T$ denoting transpose), where the superposition varies depending on the electron's azimuthal coordinate $\phi$. 

%% file: Sections/SectionScatteringRosa.tex
\section{Scattering problem} \label{scattering}
We now formulate our scattering problem to compute the spin-resolved transmission probabilities of our system. The Aharonov-Bohm-Casher ring is coupled to two fermionic reservoirs through two junctions placed along the ring, which act as scatterers. Thus, incident electronic states enter the ring structure and they are scattered coherently in the two ring arms. Within each arm with a length $\ell_i$ ($i\in 1,2$ with $\delta \ell=\ell_1-\ell_2$), electrons propagate and acquire distinct dynamical phases. For electrons traversing the $\ell_1$ arm, the accumulated phase is $k_1(\pi + \theta)$, whereas propagation through the $\ell_2$ arm leads to a phase of $k_2(\pi - \theta)$. Here, $\theta = \frac{\delta \ell}{2\ell_2 + \delta \ell} \pi$ accounts for the angular asymmetry of the junctions within the ring. 

By linearizing around the Fermi energy $\mu$, and accounting for both Fermi wavevector renormalization ($k_\mu \rightarrow \tilde{k}_\mu$) and potential shifts introduced via the gate voltage in the SOI-free arm, the respective wave numbers read
\begin{gather*}
 k_1=[\tilde{k}_\mu+(E-\mu)/\hbar v_d]b, \\ k_2=[\tilde{k}_\mu+[E-(\mu+eV_g)]/\hbar v_d]b.
\end{gather*}
 
To parametrize position within the ring and lead system, two local coordinates are introduced, as shown in Fig.~\hyperref[fig:ABinter]{1(b)}. On the left side, position is described by $x$, with $x=0$ located at the left junction where the left scatterer is placed. The electronic wavefunctions in this region are expanded in terms of plane waves $\exp(\pm ik_x x)$: positive wavevectors represent waves incident towards the ring, while negative ones correspond to reflected (outgoing) states. On the right side, the coordinate $z$ is defined such that $z=0$ marks the right junction scatterer. Here, the basis $\exp(\pm ik_z z)$ is used, but with the reversed convention: positive $k_z$ denotes outgoing waves (away from the ring), whereas negative $k_z$ describes incoming waves (entering the ring).

The asymmetric placement of the scatterers in the contact-ring system breaks rotational symmetry, so that a separate treatment for upper and lower ring segments is required. The upper arc is parameterized by an angular coordinate $\phi \in [0, \pi+\theta]$ (measured anticlockwise), where $\phi = 0$ corresponds to the right junction and $\phi = \pi+\theta$ to the left. Conversely, the lower arc is described via $\phi' \in [0, \pi-\theta]$, also increasing anticlockwise, with $\phi' = 0$ at the left junction and $\phi' = \pi-\theta$ at the right. Within the scattering-matrix formalism, one must determine the relationships between incoming and outgoing amplitudes at each junction. This involves expressing the wavefunction in each arm in terms of its local eigenstates. For the upper (SOI-active) arc, the wavefunction is expanded as
\begin{equation}
    \psi_{\ell_1} = e^{-i(\Phi/\Phi_0)\phi} \sum_{j = 1,2} \sum_{\kappa = \pm} e^{i(\Phi_{AC}^{(j)}/\Phi_0)\phi} c_{\kappa}^{(j)} e^{i \kappa k_2 \phi} \chi^{(j)},
    \label{BottomBasis}
\end{equation}
where the spinors $\chi^{(j)}$ ($j = 1,2$) are the eigenvectors of the combined Aharonov-Bohm-Casher Hamiltonian, given by Eq. \eqref{eq:2.8}. The index $j$ accounts for the two spin-split transport channels arising from Rashba SOI, and $\kappa = \pm$ labels the propagation direction, i.e., $+$ for anticlockwise, $-$ for clockwise.

By contrast, in the lower arc (where only the Aharonov-Bohm phase is relevant), the wavefunction takes the form
\begin{equation}
    \psi_{\ell_2} = e^{-i(\Phi/\Phi_0)\phi'} \sum_{\kappa = \pm} d_{\kappa} e^{i \kappa k_2 \phi'} 
    \label{BottomBasis}
\end{equation}

To determine the reflection and transmission coefficients, we adopt the Landauer-Büttiker approach, by which the incoming and outgoing wave amplitudes at each junction are related through a linear relation by the scattering matrix. Due to the presence of spin-orbit interaction, the spin-up and spin-down components mix during propagation, so they must be treated independently. Accordingly, we introduce the complete spinor state vector
\[
\Psi_{in/out} = \begin{pmatrix}
    \psi_{Lead}, &
    \psi_{\ell_1}, &
    \psi_{\ell_2}
\end{pmatrix}_{in/out}
\]
with each $\psi_q = (\psi_q^\uparrow, \psi_q^\downarrow)$ for $q \in \{Lead, \ell_1, \ell_2\}$. The scattering matrix then relates incoming and outgoing wave amplitudes as $\Psi_{out}^T = S \Psi_{in}^T$.

{The scattering matrix can be written in block form as \cite{Moskalets2011}
\begin{equation}
    {\mathcal{S}} =
    M \otimes I_2, \quad
    M = \begin{pmatrix}
        -\sqrt{1-2\epsilon} & \sqrt{\epsilon} & \sqrt{\epsilon} \\
        \sqrt{\epsilon} & a_{-} & a_{+} \\
        \sqrt{\epsilon} & a_{+} & a_{-}
    \end{pmatrix},
\end{equation}
with $I_2$ the identity matrix of dimension 2, $a_{\pm} = \frac{1}{2}(\sqrt{1-2\epsilon} \pm 1)$, and $\epsilon \in [0,0.5]$ represents the coupling strength of the junctions.}

For the left junction, we consider an ingoing wave of amplitude 1. In this case, the wavefunctions in $\ell_1$-arc is evaluated at $\phi = \pi + \theta$, and those in $\ell_2$-arc at $\phi' = 0$. In the right lead, only transmitted (outgoing) waves exist -- with spin amplitudes $t_\uparrow$ and $t_\downarrow$. Here, $\ell_1$ wavefunctions are evaluated at $\phi = 0$ and $\ell_2$ at $\phi' = \pi - \theta$. Thus, for the right junction, the incoming state is $\Psi_{in} = (0, 0, \psi_{\ell_1}^{\uparrow}, \psi_{\ell_1}^{\downarrow}, \psi_{\ell_2}^{\uparrow},  \psi_{\ell_2}^{\downarrow})_{in}$
and the outgoing state is given by $\Psi_{out} =  ( t_{\uparrow}, t_{\downarrow}, \psi_{\ell_1}^{\uparrow},  \psi_{\ell_1}^{\downarrow},  \psi_{\ell_2}^{\uparrow}, \psi_{\ell_2}^{\downarrow})_{out}$.
Finally, by solving $\Psi_{out}^T = S\Psi_{in}^T$ at the appropriate angular positions ($x=0$ and $z=0$), the spin-resolved transmission probability for spin $\sigma \in \{\uparrow,\downarrow\}$ follows as
\begin{equation}
    \mathcal{T}_{\sigma}(E) = \sum_{\sigma' \in \{\uparrow,\downarrow\}} t_{\sigma \sigma'}^\dagger(E) t_{\sigma \sigma'}(E), \label{transmissionSOI}
\end{equation}
where $\sigma'$ and $\sigma$ label the spin states of the incoming and outgoing channels, respectively.

%% file: Sections/SectionTransportRosa.tex
\section{Thermoelectric transport in the linear regime} \label{transport}

We are interested in studying charge and heat transport in the linear regime, when a small voltage bias ($\Delta V$) and a temperature gradient ($\Delta T$) are applied between the left and right reservoirs. For that purpose we employ the Landauer-Büttiker formalism to obtain the conductance matrix that characterizes the  charge ($I$) and heat ($J$) currents 
\begin{gather*}
    I = G\Delta V + L\Delta T, \\
    J = LT \Delta V + K\Delta T,
\end{gather*}
\noindent
where $G$ is the electric conductance, $L$ the thermoelectric conductance, $K$ the thermal conductivity, and $T$ the background temperature of the system. All these conductances can be determined by the following integral
\begin{equation}
    I_\mathcal{P}^{(n)} = \int_{-\infty}^\infty \mathcal{T}_\mathcal{P}(E)(E-\mu)^n\left(-\frac{\partial f(E)}{\partial E}\right)dE, \label{eq:17}
\end{equation}
\noindent
where $\mathcal{T}_\mathcal{P}$ represents the transmission probability of the transport channel $\mathcal{P}$, $\mu$ the Fermi energy, and $f(E)$ is the equilibrium Fermi-Dirac distribution. 

The presence of Rashba SOI breaks spin-rotational symmetry, rendering the projection of spin along the $z$-axis no longer a good quantum number. Consequently, the transmission for each spin channel $\mathcal{P}=\sigma$ includes spin-mixing terms, i.e., transitions between $\uparrow$ and $\downarrow$ states, as described in Eq.~(\ref{transmissionSOI}). Then, the thermoelectric coefficients in terms of the integral given by Eq. \eqref{eq:17} are given by
\begin{equation} 
G = \frac{e^2}{h}\sum_\mathcal{P} I_\mathcal{P}^{(0)}, \quad  L = \frac{e}{hT} \sum_\mathcal{P} I_\mathcal{P}^{(1)}, \quad  K = \frac{1}{hT}\sum_\mathcal{P} I_\mathcal{P}^{(2)}. \label{eq:GLK}
\end{equation}

 A key parameter to characterize thermoelectric transport is the Seebeck coefficient $S$ (or thermopower), which quantifies the voltage generated across a material in response to a temperature gradient, under open-circuit conditions. A complete thermoelectric characterization further requires the thermal conductance $\kappa_{th}$, which captures the ability of a system to conduct heat in the absence of charge current. These two coefficients can be expressed as a function of those given in Eq. \eqref{eq:GLK} as
\begin{equation}
    S = \left.-\frac{\Delta V}{\Delta T}\right|_{I=0} = \frac{L}{G}, \qquad 
    \kappa_{th} = \left.-\frac{J}{\Delta T}\right|_{I=0} = K - S^2 G T.
\end{equation}

In the low-temperature limit, where the integrals appearing in Eq.~(\ref{eq:17}) can be evaluated via the Sommerfeld expansion, the Seebeck coefficient exhibits a direct proportionality to the energy derivative of the transmission function: $S \propto \partial \mathcal{T}(E)/\partial E$, a relation known as the Mott formula. Quantum confined conductors often display pronounced energy dependence in their transmission probability, so sharp features in $\mathcal{T}(E)$ can dramatically enhance thermopower. Additionally, in this regime, the Wiedemann-Franz law links the thermal and electrical conductances through a universal proportionality.

The thermoelectric figure of merit, $ZT$~\cite{singh20, Okawa_20, Khat, Kim}, is a dimensionless coefficient that measures the engine's performance. It characterizes the efficiency of purely thermoelectric heat conversion in the absence of an applied voltage. It is expressed as
\begin{equation}
    ZT = \frac{G S^2 T}{\kappa_{th}}.
\end{equation}

Higher values of $ZT$ correspond to greater thermoelectric efficiency, which is why enhancing the figure of merit is key for optimizing the performance of quantum heat engines~\cite{Bereta, Rafael}.

%% file: Sections/SectionResultsRosa.tex
\section{Results} \label{sec:results}
In this section, we present the results for the thermoelectric behavior of the Aharonov-Bohm-Casher ring illustrated in Fig.~\ref{fig:ABinter}. We will consider a patterned GaAs ring (effective mass $m^*=0.067 m_e$, Fermi velocity $v_d=10^6$ m/s, with Rashba strength values $\alpha_R \approx  (1-5)\times 10^{-13}$ $\text{eV}\cdot \text{m}$). Unless stated otherwise, the system parameters used are: $L \equiv \ell_2 = 5\: \mu m$ for the bottom arm length, $\delta \ell =0.3\ell_2$ for the length asymmetry,  $\epsilon=0.2$ to characterize the T-junction transmissivity, $T=0.5 \: \mathrm{K}$, and $eV_g=\pi\mu$ with $\mu$ as the Fermi energy. We also set $\tilde{k}_{\mu}=\sqrt{\pi}{k}_{\mu}$ as the energy offset. 
For practical purposes, we redefine the Rashba strength such that $\tilde{\eta}=\eta -1$.  Thus, $\tilde{\eta}=0$ indicates no Rashba SOI.

\begin{figure*}[]
    \centering
    \begin{minipage}{0.33\textwidth}
        \begin{overpic}[width=\linewidth]{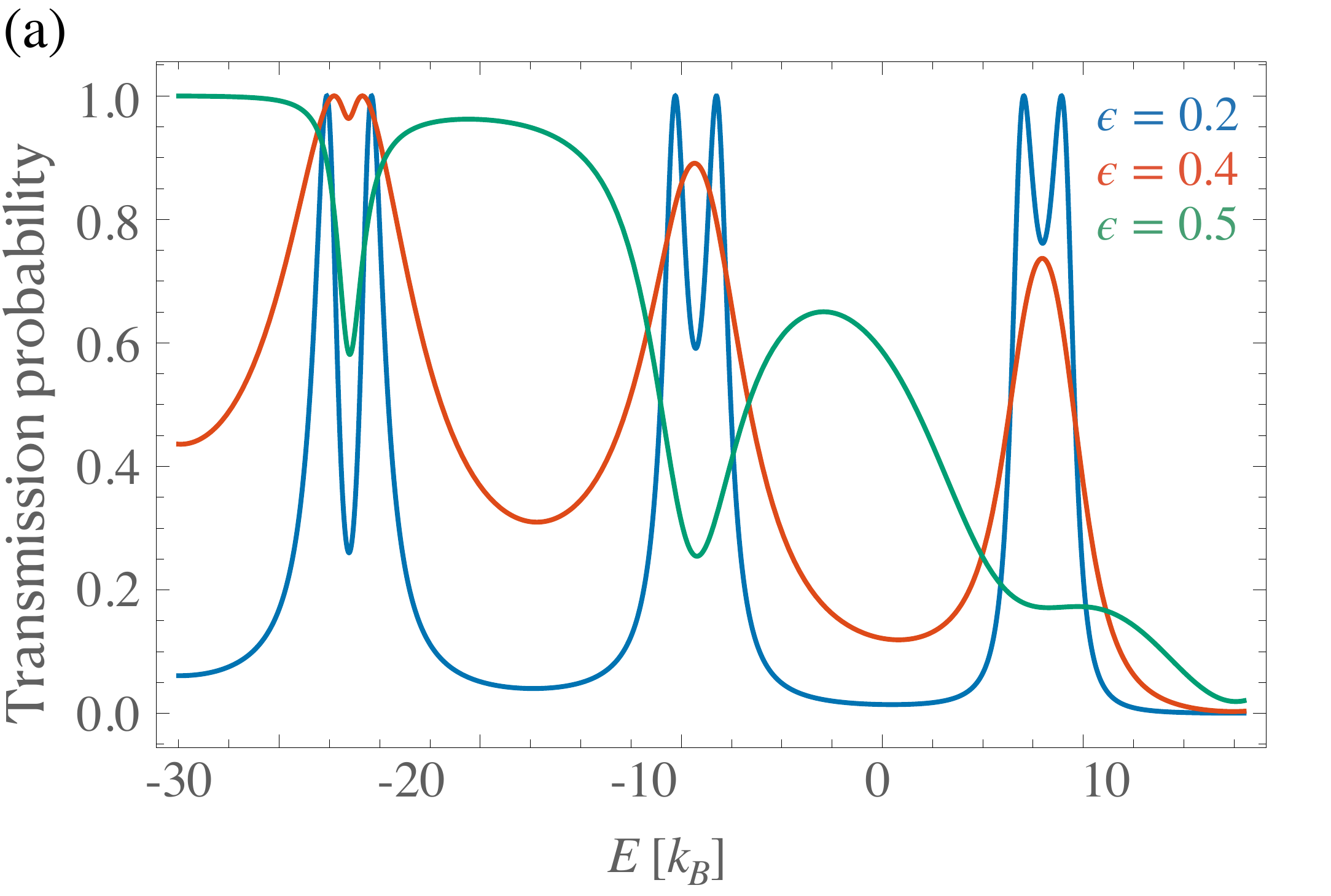}
        \end{overpic}
        \label{partialTe}
    \end{minipage}%
    \hfill
    \begin{minipage}{0.33\textwidth}
        \begin{overpic}[width=\linewidth]{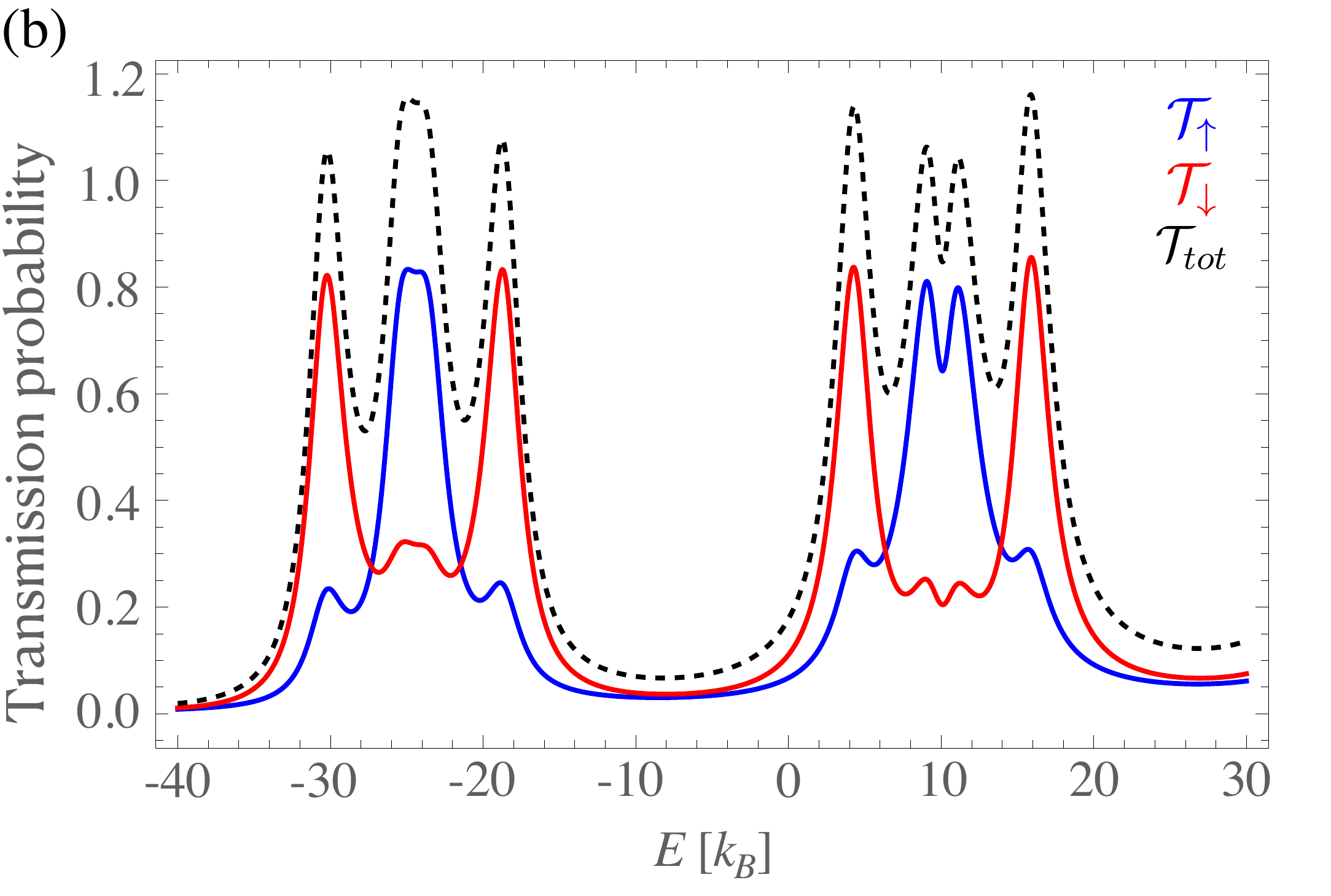}
        \end{overpic}
        \label{partialT1}
    \end{minipage}
    \hfill
    \begin{minipage}{0.33\textwidth}
        \begin{overpic}[width=\linewidth]{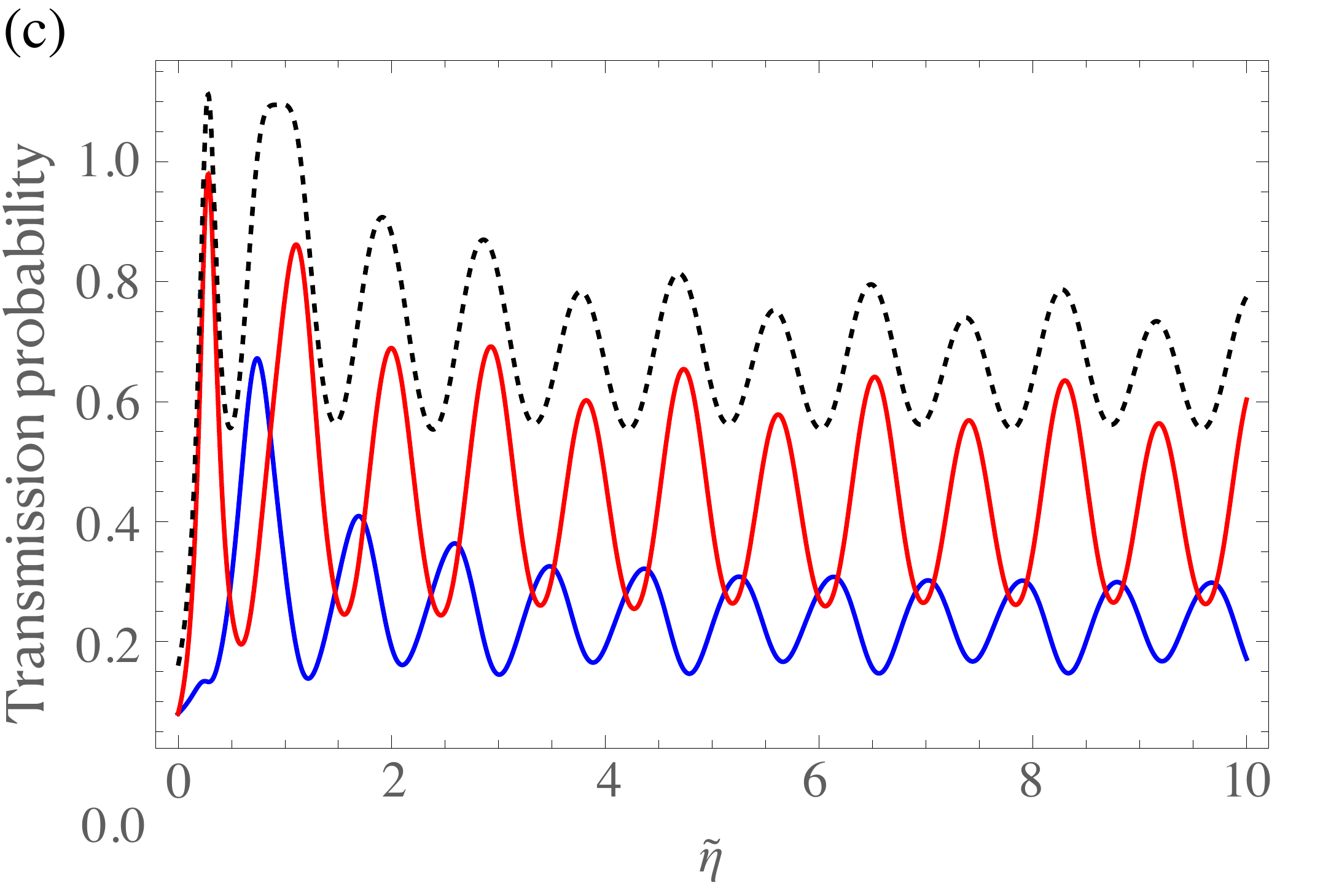}
        \end{overpic}
        \label{partialT2}
    \end{minipage}
    \caption{Transmission probability of the quantum ring. (a) Energy dependence of the transmission in absence of Rashba SOI for different coupling strength $\epsilon$, for $eV_g = \pi \mu$, and $\Phi = 1/14 \Phi_0$. (b) Energy dependence of the spin-resolved transmission probabilities ($\mathcal{T}_\uparrow, \mathcal{T}_\downarrow$) and total transmission ($\mathcal{T}_{tot}$) for $\tilde{\eta} = 0.1$ and $\epsilon = 0.2$. (c) Same quantities as in (b) as a function of $\tilde{\eta}$ at fixed energy $E = \mu$. In (a), both channels are degenerate and the plotted transmission correspond to a single channel. In (b) and (c), the total transmission can exceed unity since the contributions of both channels are shown separately.}
    \label{fig:partialTss}
\end{figure*}

Let us focus on the behavior of the transmission coefficient $\mathcal{T}(E)$ as a function of energy $E$, which determines the main features of the conductances, i.e., the  thermoelectric response of our system. Fig.~\hyperref[fig:partialTss]{2(a)} illustrates the simplest configuration:  a magnetic field threading the ring in the absence of the Rashba SOI ($\tilde{\eta}=0$). Here, the transmission $\mathcal{T}(E)$ versus the energy displays a series of resonances which broaden when the T-junction transmissivity $\epsilon$ enhances ($\epsilon=0.2\rightarrow 0.4\rightarrow 0.5$). These resonances can be easily explained by the \textit{dwell time} ($\tau_d$) \cite{Datta}, which can be defined as the average time an electron spends inside a mesoscopic system, and is related to the resonance width $\Gamma$ as 
\begin{equation*}
\tau_d = \frac{\hbar}{\Gamma}.
\end{equation*}

\noindent
This relation is understood from the energy-time uncertainty principle, since a well defined energy state (small $\Gamma$) implies a longer confinement time ($\tau_d$). When $\epsilon$ is small ($\epsilon \rightarrow 0$), electrons spend more time traveling inside the ring, resulting in a longer dwell time. As a consequence, the observed resonance peaks are narrower. On the other hand, for a strong coupling ($\epsilon \approx 0.5$), the resonances are much broader, as electrons can escape the ring much faster.

In the presence of the Rashba SOI, the spin degeneracy is lifted, and new spin-dependent phases arise. This is nicely illustrated in the spin-resolved transmission, i.e.,  $\mathcal{T}_\uparrow$ and $\mathcal{T}_\downarrow$ displayed in Fig.~\hyperref[fig:partialTss]{2(b)}. Due to the different spin-dependent phases, interference conditions are modified. This leads to spin-resolved transmission probabilities that are oscillatory and out of phase, producing a series of peaks that shift and split for each spin channel, unlike the degenerate case with no SOI. The behavior of the spin resolved transmission when the Rashba SOI is tuned is illustrated in Fig.~\hyperref[fig:partialTss]{2(c)}. Importantly, we observe two main features, namely (i) Rashba SOI strength modulates the spin-dependent transmission, as both spin components oscillate with $\tilde\eta$. Such oscillations become regular for moderate Rashba SOI fields. (ii) In consequence, the Aharonov-Bohm-Casher ring acts as a spin polarizer device in which $\mathcal{T}_\uparrow-\mathcal{T}_\downarrow\neq 0$, yielding net spin currents with nonpolarized injected electrons from the reservoirs. Besides, the spin polarization maximizes for relatively large Rashba SOI intensities when both spin transmission signals are out of phase.
  
\begin{figure*}[]

    \centering

    \begin{minipage}{0.32\textwidth}
        \begin{overpic}[width=\linewidth]{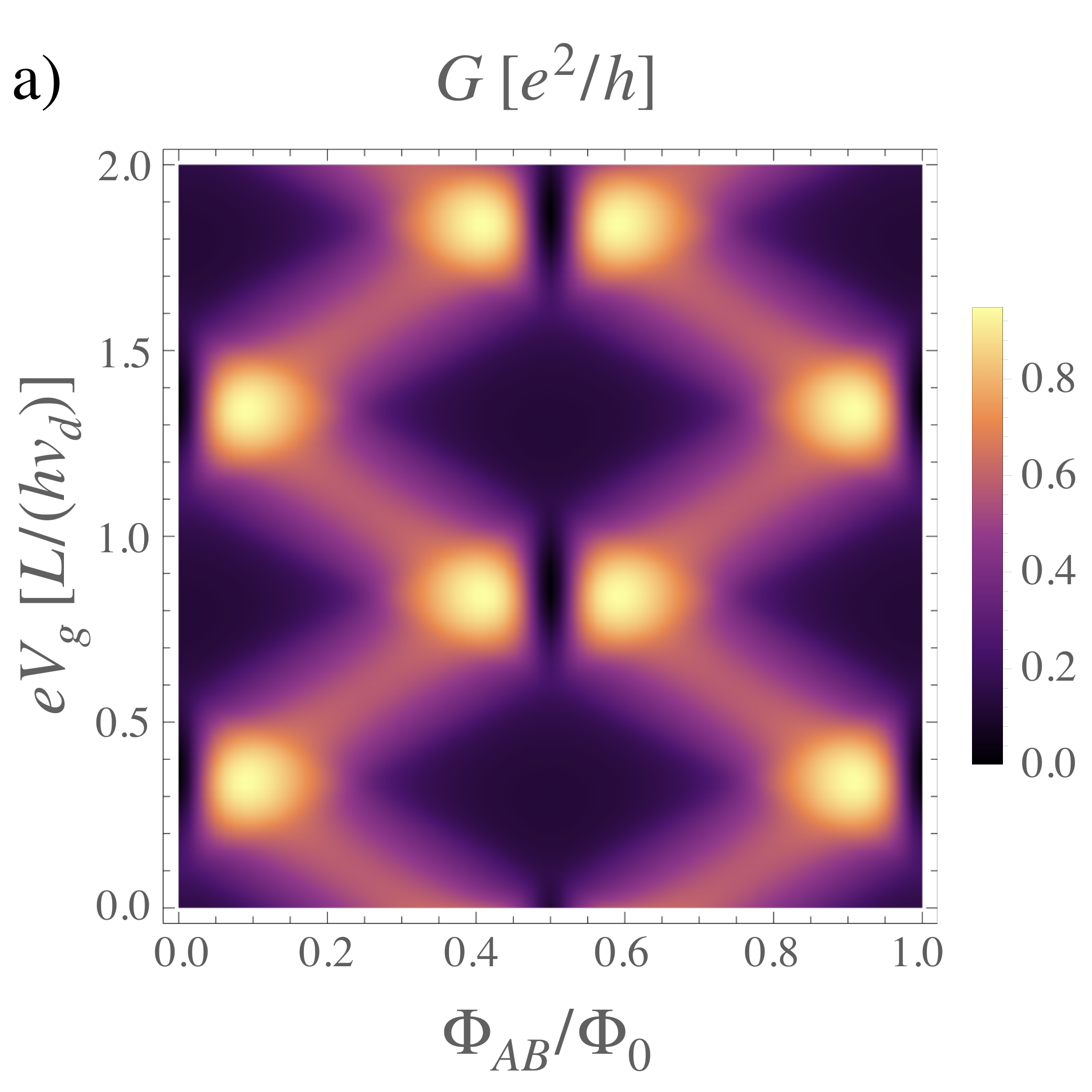}
        \end{overpic}
        \label{fig:spectrumAB}
    \end{minipage}%
    \hfill
    \begin{minipage}{0.32\textwidth}
        \begin{overpic}[width=\linewidth]{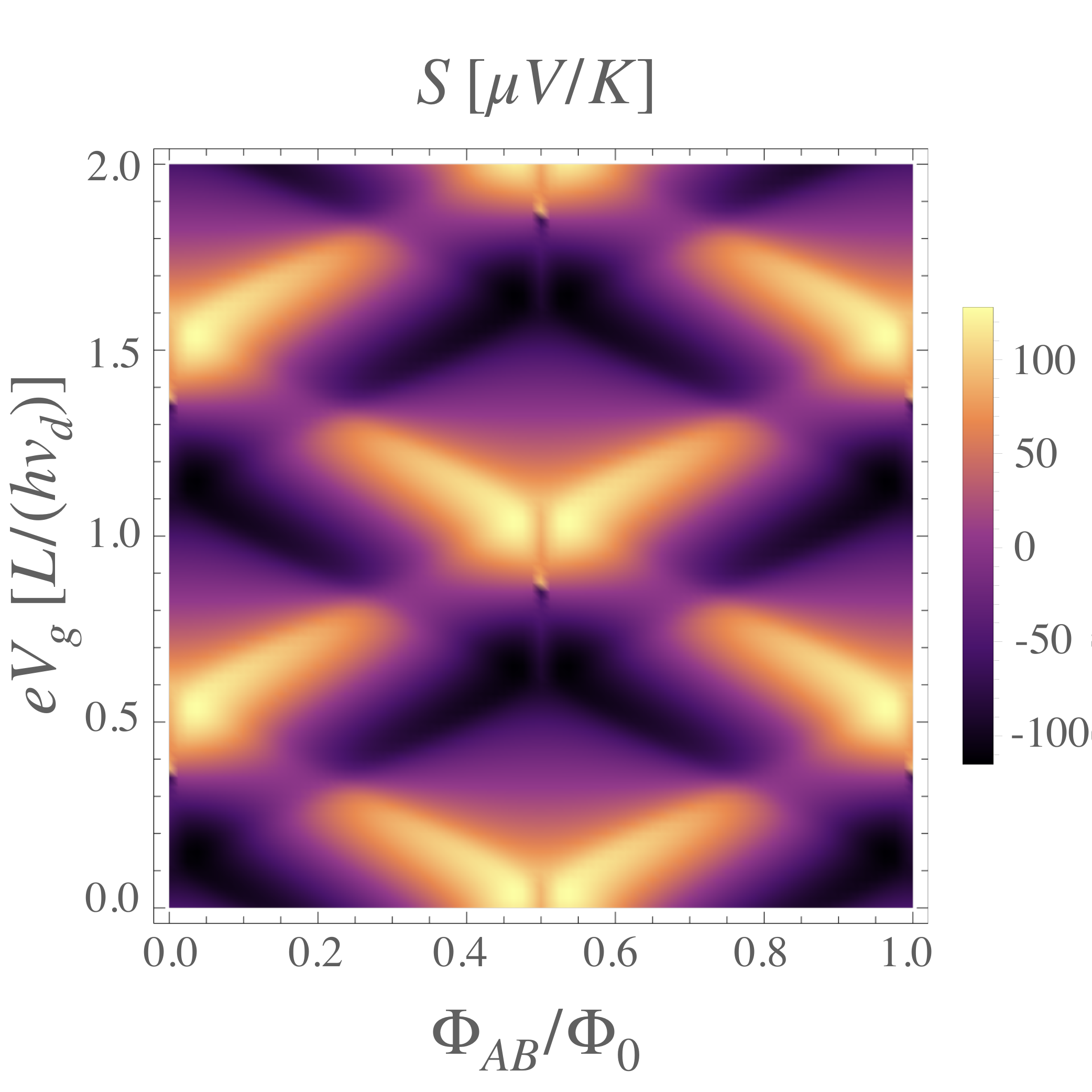}
        \end{overpic}
        \label{spectrumAB2}
    \end{minipage}%
    \hfill
    \begin{minipage}{0.32\textwidth}
        \begin{overpic}[width=\linewidth]{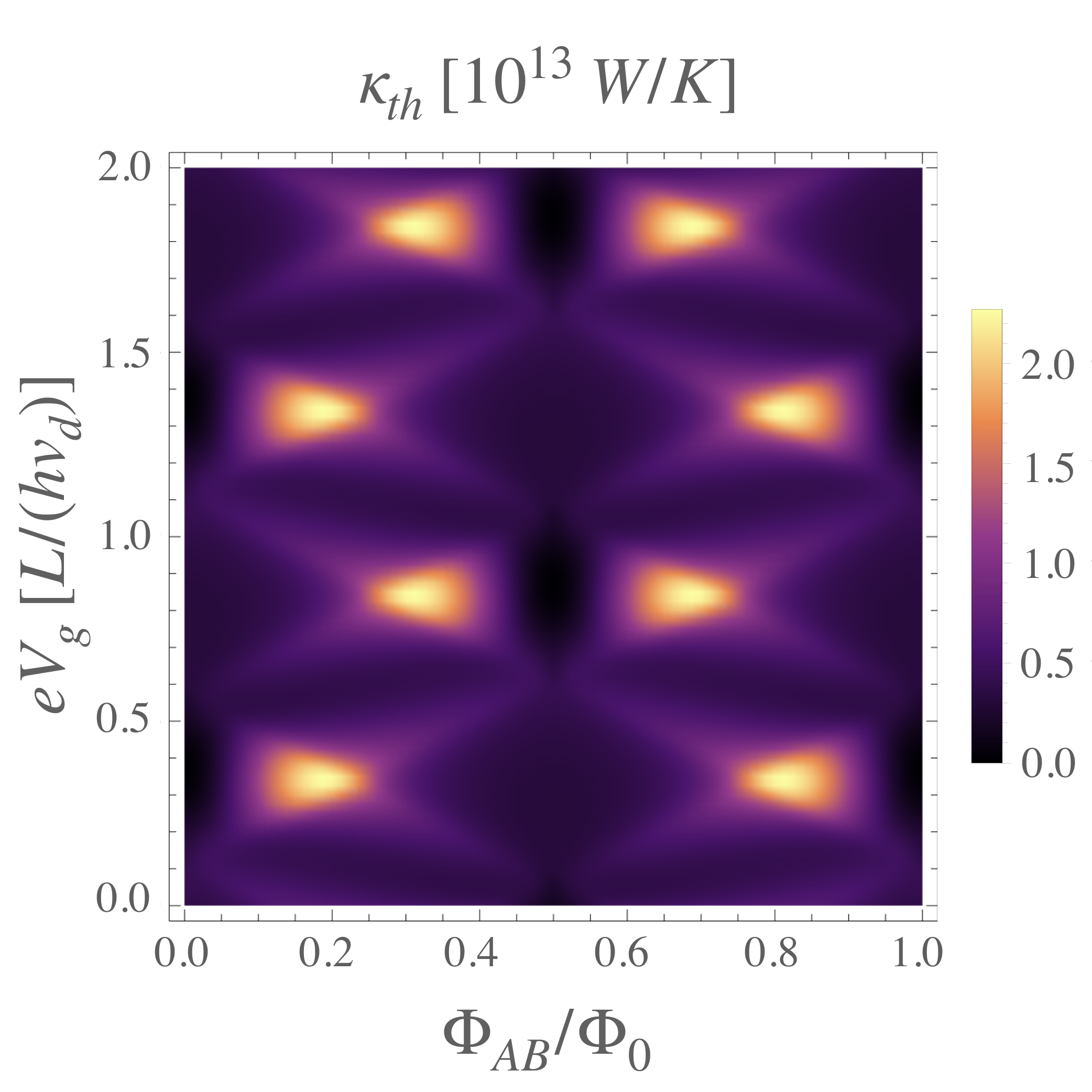}
        \end{overpic}
        \label{spectrumAB3}
    \end{minipage}

    \begin{minipage}{0.32\textwidth}
        \begin{overpic}[width=\linewidth]{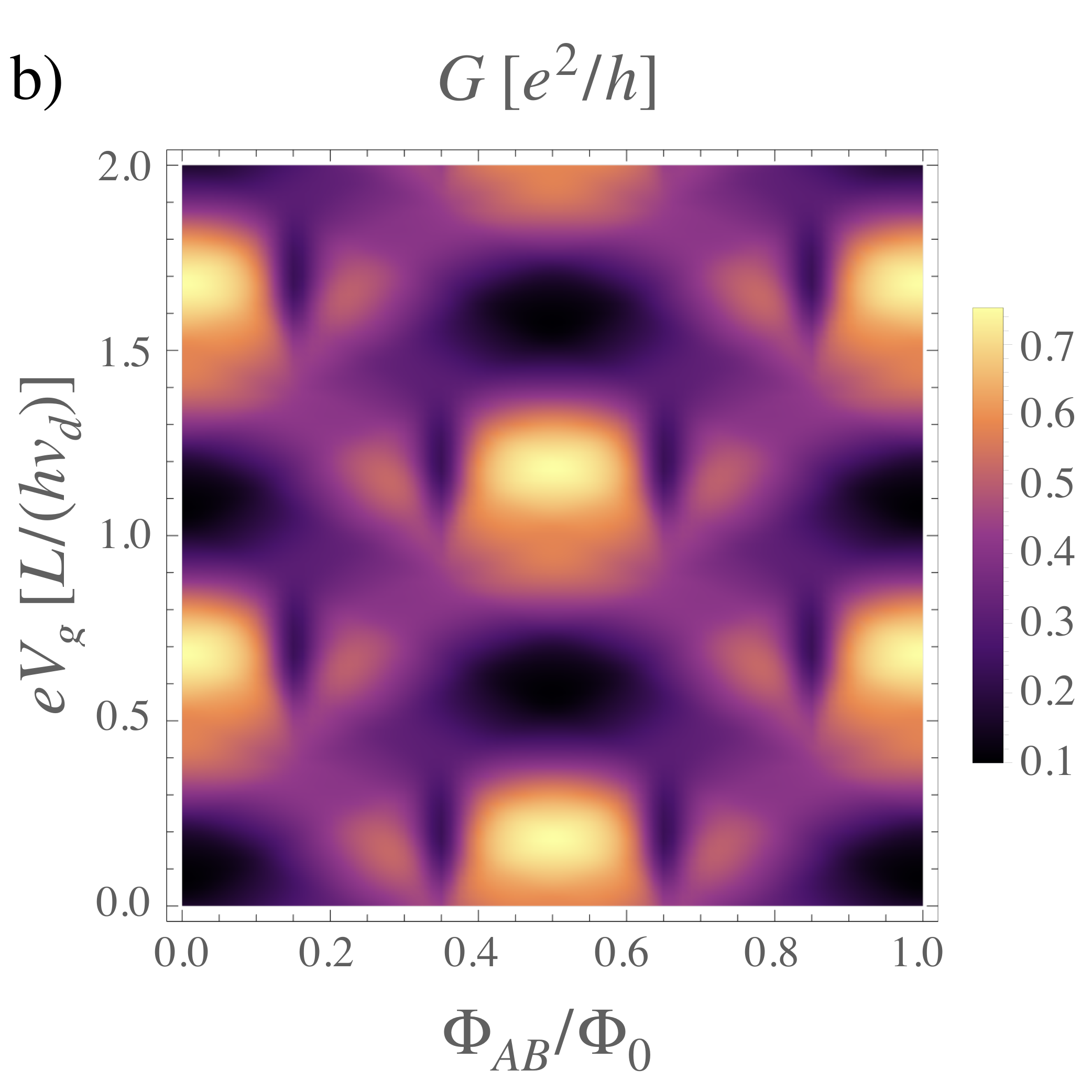}
        \end{overpic}
        \label{spectrumAB4}
    \end{minipage}%
    \hfill
    \begin{minipage}{0.32\textwidth}
        \begin{overpic}[width=\linewidth]{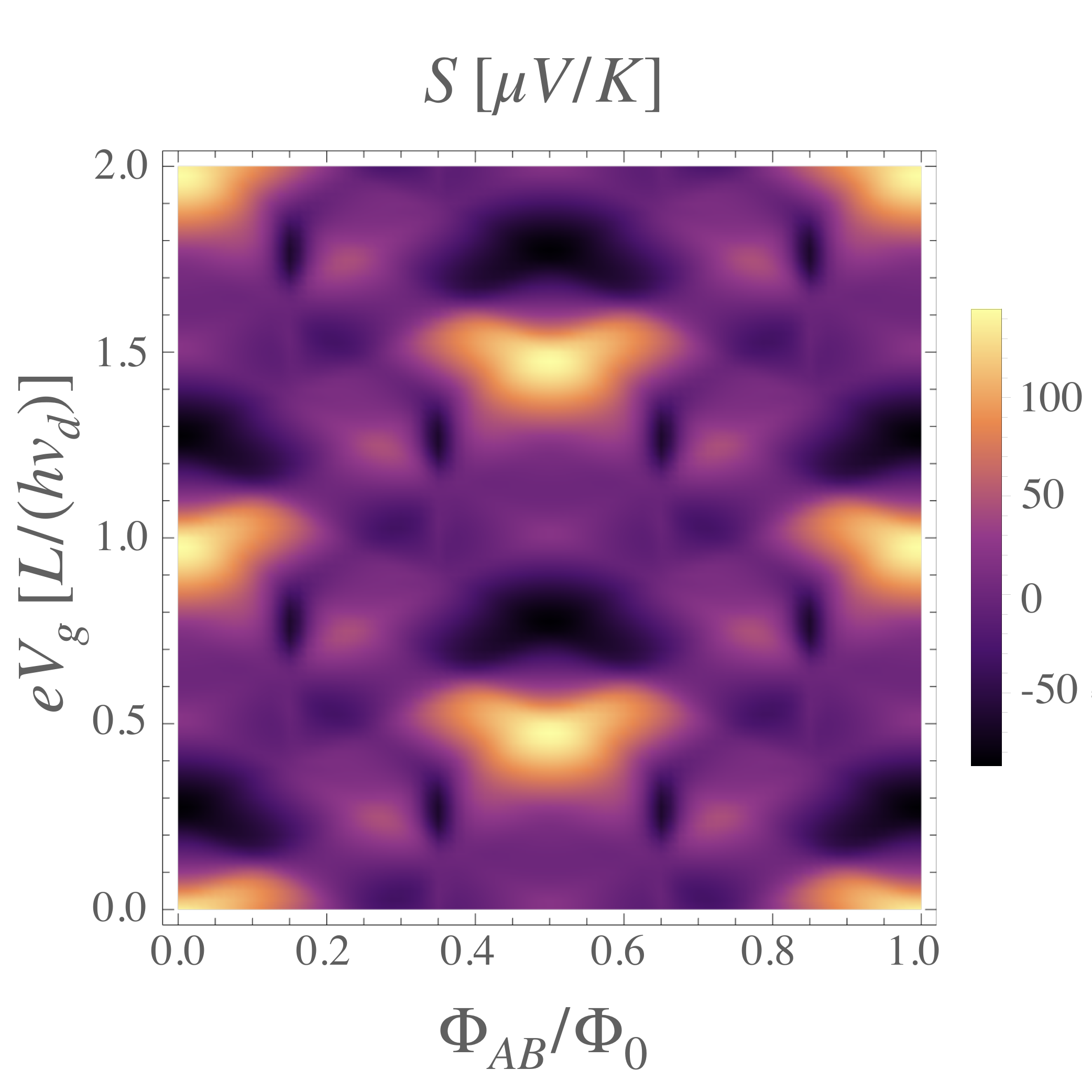}
        \end{overpic}
        \label{spectrumAB5}
    \end{minipage}%
    \hfill
    \begin{minipage}{0.32\textwidth}
        \begin{overpic}[width=\linewidth]{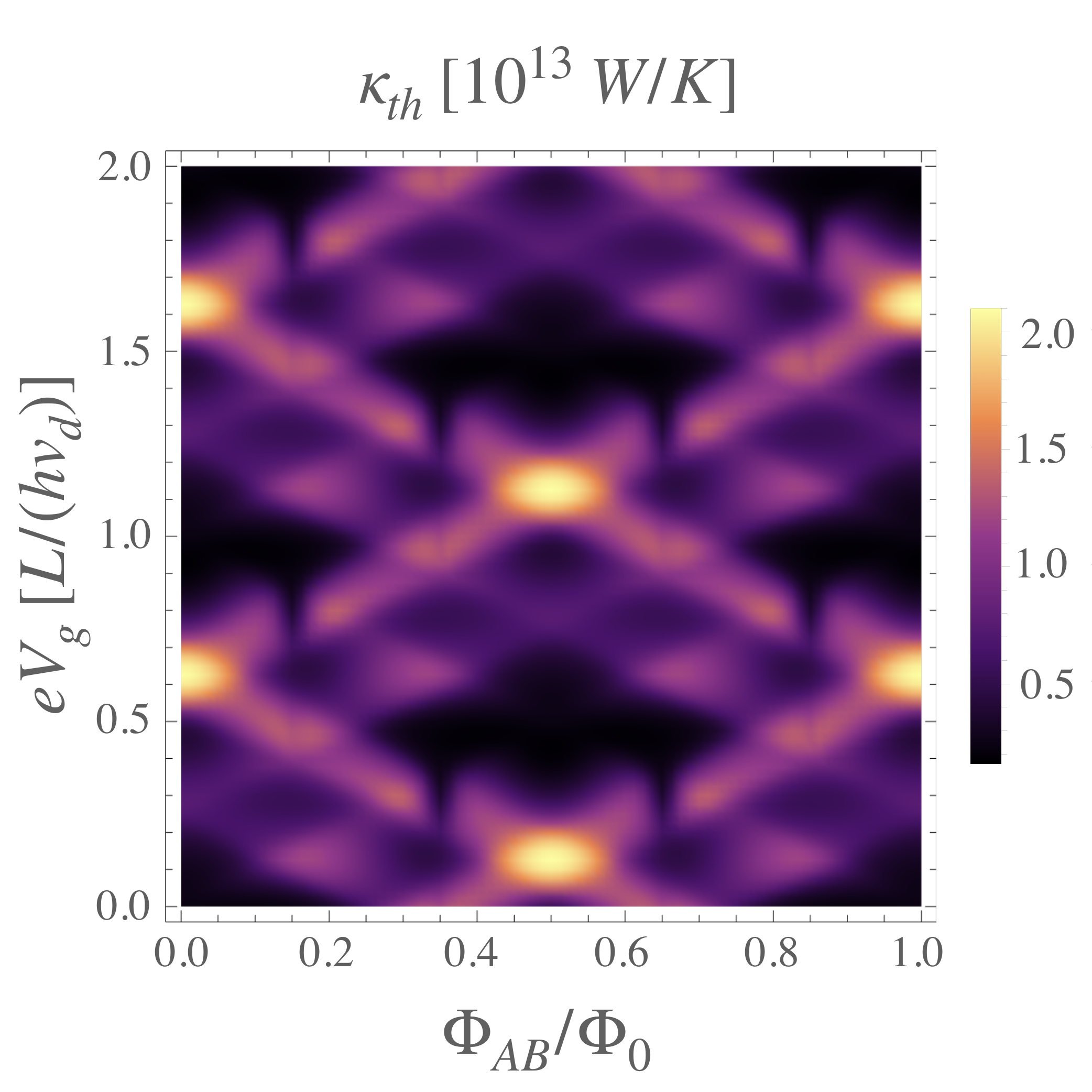}
        \end{overpic}
        \label{spectrumAB6}
    \end{minipage}
    \vspace{-0.1 cm}
    \caption{Thermoelectric behavior of the Aharonov-Bohm ring-based quantum heat engine. (a) Contour plots of the normalized conductance $G$, charge Seebeck $S$ and thermal conductance  $\kappa_{th}$ for the case of no Rashba SOI ($\tilde{\eta} = 0$) vs gate voltage $V_g$ and magnetic flux $\Phi_{AB}$. (b) Same quantities in the presence of Rashba spin-orbit coupling with $\tilde{\eta} = 0.2$.}
    \label{fig:coefficients}
\end{figure*}

The transmission characteristics determine the thermoelectric behavior of the Aharonov-Bohm-Casher ring. Our study shows the behavior of the electric conductance $G$, the charge Seebeck $S$ and the thermal conductance $\kappa_{th}$ as a function of gate voltage and magnetic flux for different system parameters as the arm asymmetry and the Rashba SOI strength. Fig.~\hyperref[fig:coefficients]{3} shows contour maps for these coefficients versus the gate voltage $eV_g$ and magnetic flux $\frac{\Phi_{AB}}{\Phi_0}$. The upper panel corresponds to the ring configuration without Rashba SOI, whereas the lower panel shows the case where Rashba strength $\tilde\eta =0.2$. Both cases are displayed for an asymmetric ring, i.e., $\delta \ell=0.3 \ell_2$.

In the absence of Rashba SOI (see Fig.~\hyperref[fig:coefficients]{3(a)}), all transport features ($G$, $S$, and $\kappa_{th}$) exhibit highlighted ridges that represent the regions where constructive interferences take place. These correspond to transmission peaks for $G$ and $\kappa_{th}$, or regions where  these transmission resonances exhibit a large electron-hole asymmetry, which enhances the thermopower. Focusing on the charge Seebeck $S$, the areas where it changes its sign correspond to those where transport switches from electron to hole-like mode. As mentioned, the Seebeck coefficient acquires higher values whenever the transmission asymmetry is higher.

Specifically, the thermal conductance displays bright and dark areas that can be controlled via the AB phase. Thus, by varying the AB flux, the thermal conductance can be switched between fully insulating and fully conducting states. In the low temperature limit (not shown here), both $G$ and $\kappa_{th}\approx K$ are proportional to the Lorentz number, a universal constant, as the Wiedemann-Franz dictates. Deviations from such behavior are encountered for moderate temperature as used in Fig.~\hyperref[fig:coefficients]{3(a)} where $T=0.5$ K. Here,  $\kappa_{th}$ (through the coefficient $K$) is built by contributions that are away from the Fermi energy  [accounted by $(E-\mu)^2$ in Eq. (\ref{eq:GLK})]. As a result, bright and dark regions for $\kappa_{th}$ no longer coincide with the corresponding bright and dark regions for $G$, a behavior which is desirable to enhance the figure of merit $ZT$, as shown later.

The addition of Rashba SOI produces a much more complex thermoelectric behavior, as seen in Fig.~\hyperref[fig:coefficients]{3(b)}. As the spin symmetry is broken, there appear two spin-resolved interference patterns with less sharp maxima, which overlap in some regions. The appearance of more interference peaks leads to richer patterns of $G$, $S$ and $\kappa_{th}$. The additional asymmetry arises from the spin-dependent phase. Notice that now spin $\sigma$ electrons feel an Aharonov-Bohm-Casher phase $\Phi_{\sigma}\rightarrow \Phi_{AB}/\Phi_0+\sigma \eta$ when they travel through the Rashba arm. The spin asymmetry
introduced with Rashba SOI leads to different interference conditions for spin-up and spin-down electrons, giving rise to new behavior for the electrical and thermal conductances. The previous bright-dark regions have split and they move apart in the non-zero Rashba configuration. Besides, the additional asymmetry introduced by the spin-dependent phases leads to a slightly higher values of the Seebeck, as displayed in Fig.~\hyperref[fig:coefficients]{3(b)}. 
All these novel features attributed to the Rashba SOI work together to achieve extraordinary large values for the ZT coefficient, the figure of merit, which quantifies how good is a device as a thermoelectric transducer, i.e., how good a device converts heat into electricity and vice versa. 
In this sense,  Rashba SOI introduces a new degree of tunability to the device that can be used to optimize thermoelectric performance through an electric field.

\begin{figure*}[t]
    \centering
    \begin{minipage}{0.32\textwidth}
        \begin{overpic}[width=\linewidth]{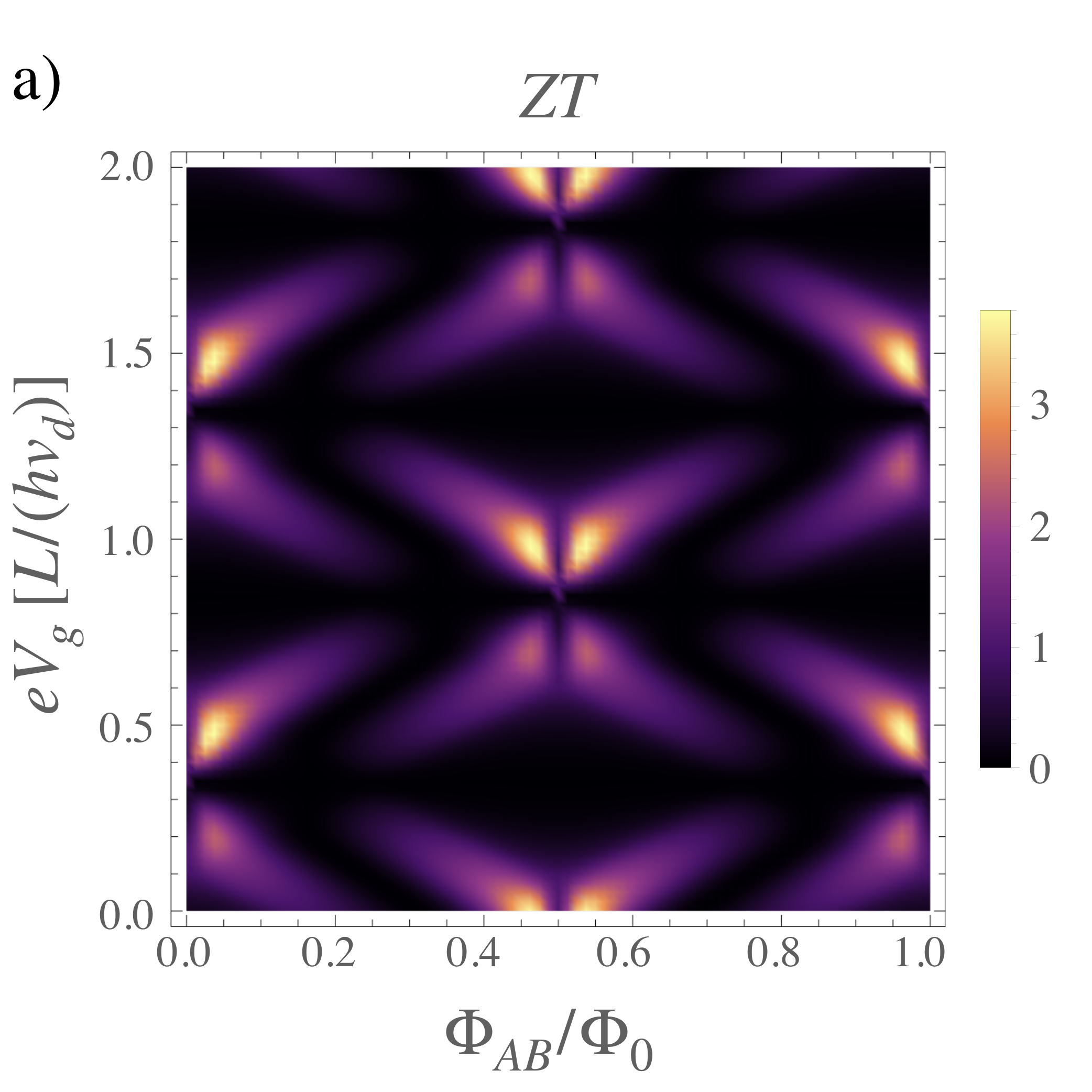}
        \end{overpic}
        \label{partialTe}
    \end{minipage}%
    \hfill
    \begin{minipage}{0.32\textwidth}
        \begin{overpic}[width=\linewidth]{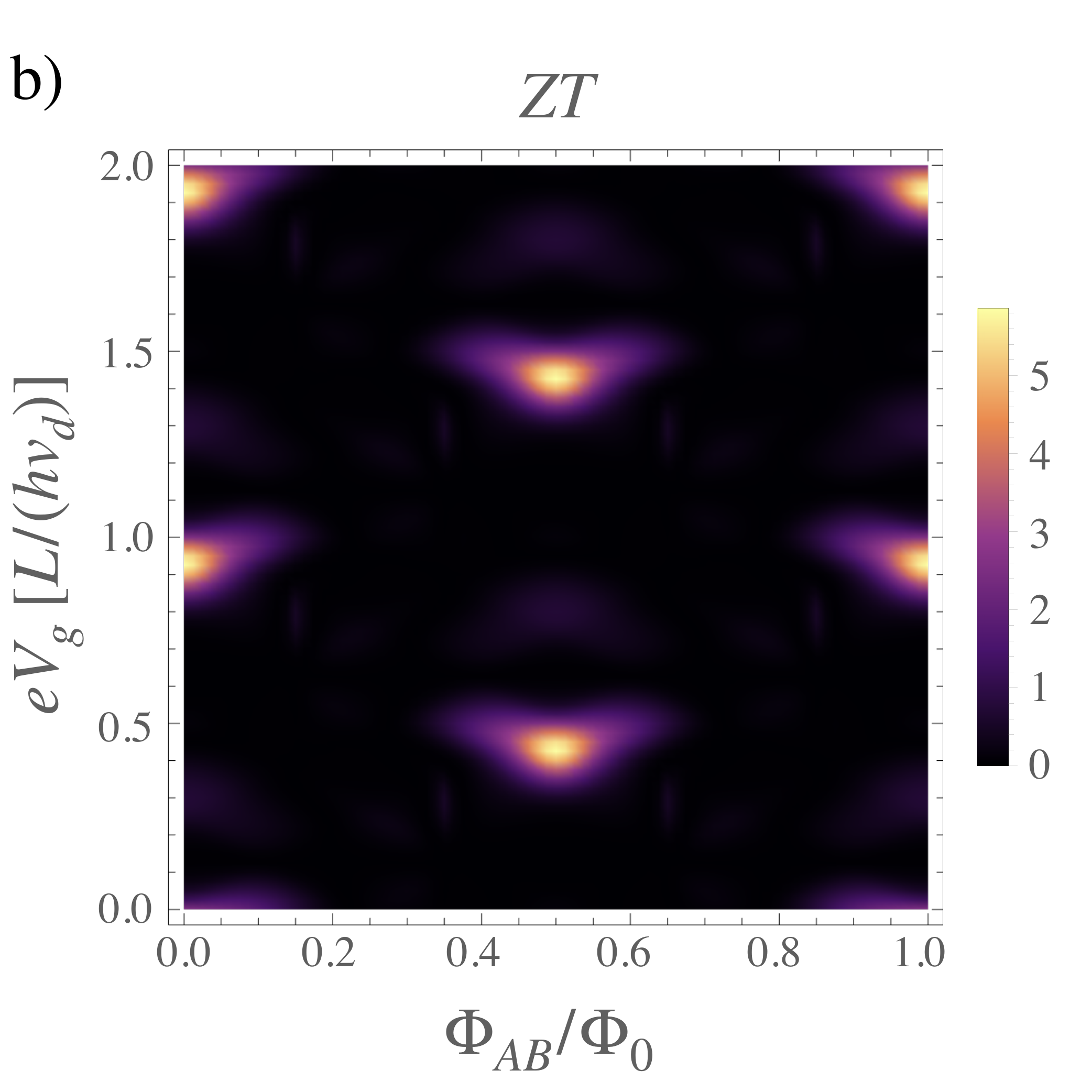}
        \end{overpic}
        \label{partialT1}
    \end{minipage}
    \hfill
    \begin{minipage}{0.33\textwidth}
        \begin{overpic}[width=\linewidth]{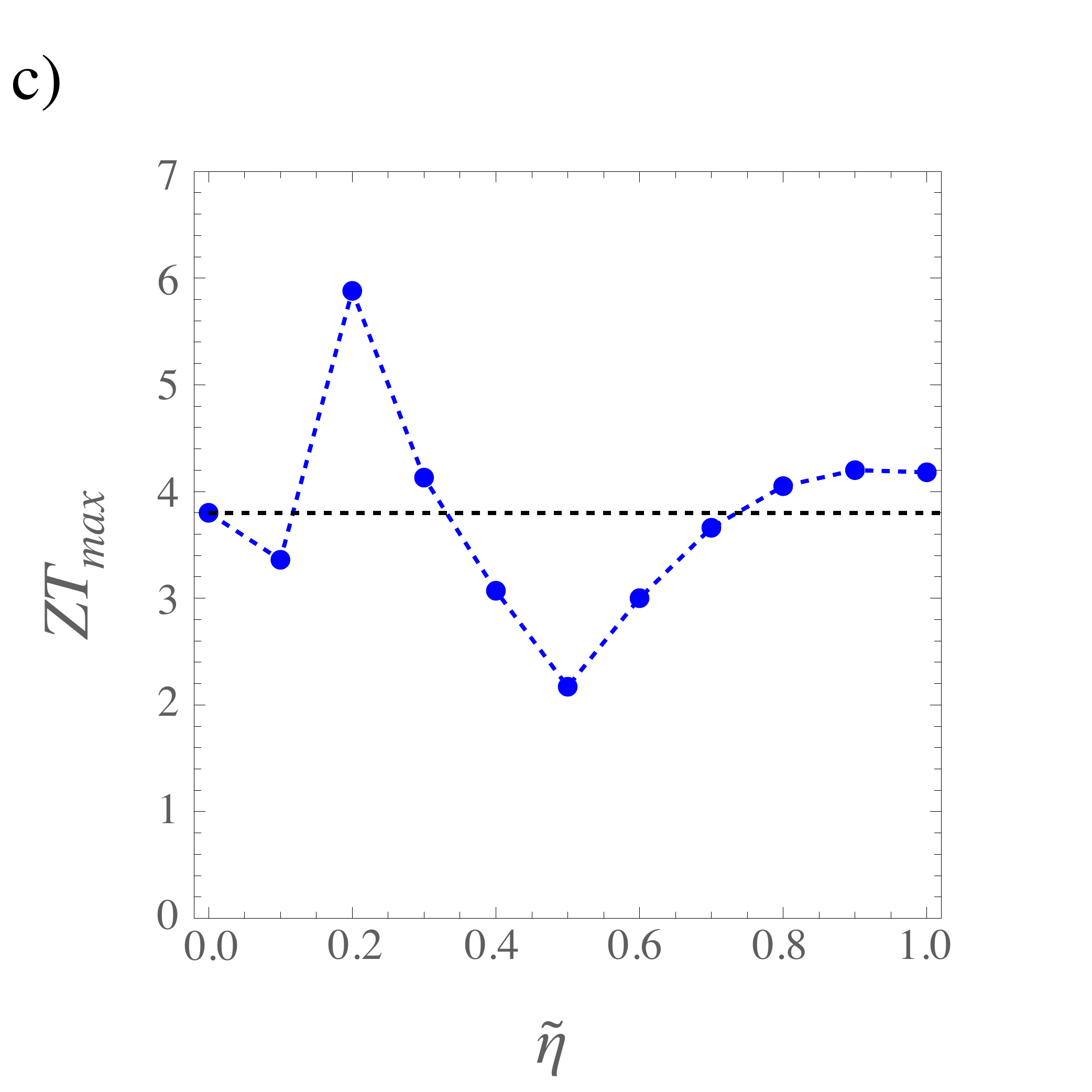}
        \end{overpic}
        \label{partialT2}
    \end{minipage}
    \vspace{-0.1cm}
    \caption{Figure of merit of the Aharonov-Bohm ring-based heat engine: (a) configuration in the absence of Rashba SOI, as shown in Fig.~\hyperref[fig:coefficients]{3(a)}, (b) configuration with $\tilde{\eta} = 0.2$, as shown in Fig.~\hyperref[fig:coefficients]{3(b)} and (c) Maximum value of $ZT$ obtained for different Rashba strengths $\tilde{\eta}$. The dotted line represents the maximum $ZT$ obtained in the absence of Rashba SOI.}
    \label{fig:zt}
\end{figure*}

The figure of merit $ZT$ is shown in Fig.~\hyperref[fig:coefficients]{4(a)} for the case with no Rashba SOI, using the same parameters as in Fig.~\hyperref[fig:coefficients]{3(a)}; and Fig.~\hyperref[fig:coefficients]{4(b)} with the presence of Rashba, using the same parameters as in Fig.~\hyperref[fig:coefficients]{3(b)}.
In both cases, the figure of merit reaches values of $ZT >1$. The regions where this occurs are those where the Seebeck $S$ is enhanced while the thermal conductance is low, reflecting an optimal balance between charge and heat transport. This is clearly seen when the regions where $ZT>1$ in Figs.~\hyperref[fig:zt]{4(a)} and \hyperref[fig:zt]{4(b)} are compared with the regions where the Seebeck coefficient is large and the thermal conductance $\kappa_{th}$ gets small, as shown in Fig.~\hyperref[fig:coefficients]{3(a)} (no Rashba SOI) and Fig.~\hyperref[fig:coefficients]{3(b)} (finite Rashba SOI).

Importantly, the largest $ZT$ value is attained for an AB ring in the presence of Rashba SOI. This is shown in Fig.~\hyperref[fig:zt]{4(c)} where the maximum value of the figure of merit, denoted by $ZT_{\rm max}$, is represented versus the Rashba strength $\tilde\eta$, maintaining the other parameters fixed. We have computed the values of $ZT$ in a region in which $eV_g\in[0,2]$ (units of $L/hv_d$, with $L\equiv \ell_2$) and $\Phi_{AB}/\Phi_0\in[0,1]$. The maximum values for the figure of merit are found always when the magnetic flux is equal to $0$ or $1/2 \Phi_0$. We predict that  $ZT_{\rm max}\approx 6$ for $\tilde{\eta} = 0.2$. This constitutes a remarkable increase of 55\% in comparison with the case of no Rashba SOI ($\tilde{\eta}=0$). Notably, our results indicate that the figure of merit $ZT$ can be tuned by electric field control, which varies the Rashba SOI strength, to reach high values.

Overall, these results show that Rashba SOI provides an efficient way to control and achieve the thermoelectric properties of the system. By properly adjusting $\Phi$ and $\tilde{\eta}$, one can tune the device to regimes of optimal performance.

%% file: Sections/SectionVIConclusionsRosa.tex
\section{Conclusions}

Our work investigates thermoelectric transport in an asymmetric Aharonov-Bohm ring where one arm incorporates Rashba spin-orbit interaction (SOI). Using a scattering matrix approach, we derived spin-resolved transmission probabilities to calculate the linear-response electrical and thermal conductances, the Seebeck coefficient, and the thermoelectric figure of merit $ZT$. By introducing Rashba SOI in a single arm and varying the magnetic flux, gate voltage, and geometric asymmetry, we demonstrate the emergence of strong spin-dependent interference patterns. At moderate Rashba strengths, the spin-resolved transmission peaks become out of phase, causing the Aharonov-Bohm-Casher ring to display novel interference features. As a consequence,  the Rashba SOI yields larger values for the Seebeck coefficient and  a more intricate behavior in both the electrical and thermal conductances. 

Finally, our calculations show that the combination of Rashba SOI and ring asymmetry, $\delta \ell\neq 0$ allows large values for the figure of merit which are significantly greater than unity, reaching $ZT_{\rm max} \approx 6$ under optimal conditions. The parameter ranges and configurations explored in this work, $\ell_2 =5\: \mu m$, $\delta \ell =0.3 \ell_2$, and $\tilde\eta=0.2$ correspond to standard experimental conditions, allowing straightforward realization and verification of the predicted effects.  Importantly, $ZT$ is nearly doubled compared to a symmetric ring, demonstrating the effectiveness of asymmetry and spin engineering for thermoelectric optimization.

In addition to the high thermoelectric efficiency obtained, our proposed setup improves the tunability of the device by introducing Rashba SOI. That way, the device can be easily configured to optimize charge and heat transport. This tunability, together with the higher values of $ZT$ obtained, offers lots of possibilities for future nanoscale device energy conversion.

Finally, while our analysis focused on the linear transport regime, our model can be extended to the nonlinear one, where stronger temperature and voltage gradients could lead to an enhanced thermoelectric performance.

%% file: mainABC.bbl
\begin{thebibliography}{26}%
\makeatletter
\providecommand \@ifxundefined [1]{%
 \@ifx{#1\undefined}
}%
\providecommand \@ifnum [1]{%
 \ifnum #1\expandafter \@firstoftwo
 \else \expandafter \@secondoftwo
 \fi
}%
\providecommand \@ifx [1]{%
 \ifx #1\expandafter \@firstoftwo
 \else \expandafter \@secondoftwo
 \fi
}%
\providecommand \natexlab [1]{#1}%
\providecommand \enquote  [1]{``#1''}%
\providecommand \bibnamefont  [1]{#1}%
\providecommand \bibfnamefont [1]{#1}%
\providecommand \citenamefont [1]{#1}%
\providecommand \href@noop [0]{\@secondoftwo}%
\providecommand \href [0]{\begingroup \@sanitize@url \@href}%
\providecommand \@href[1]{\@@startlink{#1}\@@href}%
\providecommand \@@href[1]{\endgroup#1\@@endlink}%
\providecommand \@sanitize@url [0]{\catcode `\\12\catcode `\$12\catcode `\&12\catcode `\#12\catcode `\^12\catcode `\_12\catcode `\%12\relax}%
\providecommand \@@startlink[1]{}%
\providecommand \@@endlink[0]{}%
\providecommand \url  [0]{\begingroup\@sanitize@url \@url }%
\providecommand \@url [1]{\endgroup\@href {#1}{\urlprefix }}%
\providecommand \urlprefix  [0]{URL }%
\providecommand \Eprint [0]{\href }%
\providecommand \doibase [0]{http://dx.doi.org/}%
\providecommand \selectlanguage [0]{\@gobble}%
\providecommand \bibinfo  [0]{\@secondoftwo}%
\providecommand \bibfield  [0]{\@secondoftwo}%
\providecommand \translation [1]{[#1]}%
\providecommand \BibitemOpen [0]{}%
\providecommand \bibitemStop [0]{}%
\providecommand \bibitemNoStop [0]{.\EOS\space}%
\providecommand \EOS [0]{\spacefactor3000\relax}%
\providecommand \BibitemShut  [1]{\csname bibitem#1\endcsname}%
\let\auto@bib@innerbib\@empty
\bibitem [{\citenamefont {Giazotto}\ \emph {et~al.}(2006)\citenamefont {Giazotto}, \citenamefont {Heikkil\"a}, \citenamefont {Luukanen}, \citenamefont {Savin},\ and\ \citenamefont {Pekola}}]{Giazz}%
  \BibitemOpen
  \bibfield  {author} {\bibinfo {author} {\bibfnamefont {F.}~\bibnamefont {Giazotto}}, \bibinfo {author} {\bibfnamefont {T.~T.}\ \bibnamefont {Heikkil\"a}}, \bibinfo {author} {\bibfnamefont {A.}~\bibnamefont {Luukanen}}, \bibinfo {author} {\bibfnamefont {A.~M.}\ \bibnamefont {Savin}}, \ and\ \bibinfo {author} {\bibfnamefont {J.~P.}\ \bibnamefont {Pekola}},\ }\href {\doibase 10.1103/RevModPhys.78.217} {\bibfield  {journal} {\bibinfo  {journal} {Rev. Mod. Phys.}\ }\textbf {\bibinfo {volume} {78}},\ \bibinfo {pages} {217} (\bibinfo {year} {2006})}\BibitemShut {NoStop}%
\bibitem [{\citenamefont {Benenti}\ \emph {et~al.}(2017)\citenamefont {Benenti}, \citenamefont {Casati}, \citenamefont {Saito},\ and\ \citenamefont {Whitney}}]{Benenti}%
  \BibitemOpen
  \bibfield  {author} {\bibinfo {author} {\bibfnamefont {G.}~\bibnamefont {Benenti}}, \bibinfo {author} {\bibfnamefont {G.}~\bibnamefont {Casati}}, \bibinfo {author} {\bibfnamefont {K.}~\bibnamefont {Saito}}, \ and\ \bibinfo {author} {\bibfnamefont {R.}~\bibnamefont {Whitney}},\ }\href {\doibase 10.1016/j.physrep.2017.05.008} {\bibfield  {journal} {\bibinfo  {journal} {Physics Reports}\ }\textbf {\bibinfo {volume} {694}},\ \bibinfo {pages} {1–124} (\bibinfo {year} {2017})}\BibitemShut {NoStop}%
\bibitem [{\citenamefont {Pekola}\ and\ \citenamefont {Karimi}(2021)}]{Pekola}%
  \BibitemOpen
  \bibfield  {author} {\bibinfo {author} {\bibfnamefont {J.~P.}\ \bibnamefont {Pekola}}\ and\ \bibinfo {author} {\bibfnamefont {B.}~\bibnamefont {Karimi}},\ }\href {\doibase 10.1103/RevModPhys.93.041001} {\bibfield  {journal} {\bibinfo  {journal} {Rev. Mod. Phys.}\ }\textbf {\bibinfo {volume} {93}},\ \bibinfo {pages} {041001} (\bibinfo {year} {2021})}\BibitemShut {NoStop}%
\bibitem [{\citenamefont {Potanina}\ \emph {et~al.}(2021)\citenamefont {Potanina}, \citenamefont {Flindt}, \citenamefont {Moskalets},\ and\ \citenamefont {Brandner}}]{Potanina}%
  \BibitemOpen
  \bibfield  {author} {\bibinfo {author} {\bibfnamefont {E.}~\bibnamefont {Potanina}}, \bibinfo {author} {\bibfnamefont {C.}~\bibnamefont {Flindt}}, \bibinfo {author} {\bibfnamefont {M.}~\bibnamefont {Moskalets}}, \ and\ \bibinfo {author} {\bibfnamefont {K.}~\bibnamefont {Brandner}},\ }\href {\doibase 10.1103/PhysRevX.11.021013} {\bibfield  {journal} {\bibinfo  {journal} {Phys. Rev. X}\ }\textbf {\bibinfo {volume} {11}},\ \bibinfo {pages} {021013} (\bibinfo {year} {2021})}\BibitemShut {NoStop}%
\bibitem [{\citenamefont {Sothmann}\ \emph {et~al.}(2014)\citenamefont {Sothmann}, \citenamefont {Sánchez},\ and\ \citenamefont {Jordan}}]{Sothmann}%
  \BibitemOpen
  \bibfield  {author} {\bibinfo {author} {\bibfnamefont {B.}~\bibnamefont {Sothmann}}, \bibinfo {author} {\bibfnamefont {R.}~\bibnamefont {Sánchez}}, \ and\ \bibinfo {author} {\bibfnamefont {A.~N.}\ \bibnamefont {Jordan}},\ }\href {\doibase 10.1088/0957-4484/26/3/032001} {\bibfield  {journal} {\bibinfo  {journal} {Nanotechnology}\ }\textbf {\bibinfo {volume} {26}},\ \bibinfo {pages} {032001} (\bibinfo {year} {2014})}\BibitemShut {NoStop}%
\bibitem [{\citenamefont {Blasi}\ \emph {et~al.}(2021)\citenamefont {Blasi}, \citenamefont {Taddei}, \citenamefont {Arrachea}, \citenamefont {Carrega},\ and\ \citenamefont {Braggio}}]{Blasi}%
  \BibitemOpen
  \bibfield  {author} {\bibinfo {author} {\bibfnamefont {G.}~\bibnamefont {Blasi}}, \bibinfo {author} {\bibfnamefont {F.}~\bibnamefont {Taddei}}, \bibinfo {author} {\bibfnamefont {L.}~\bibnamefont {Arrachea}}, \bibinfo {author} {\bibfnamefont {M.}~\bibnamefont {Carrega}}, \ and\ \bibinfo {author} {\bibfnamefont {A.}~\bibnamefont {Braggio}},\ }\href {\doibase 10.1103/PhysRevB.103.235434} {\bibfield  {journal} {\bibinfo  {journal} {Phys. Rev. B}\ }\textbf {\bibinfo {volume} {103}},\ \bibinfo {pages} {235434} (\bibinfo {year} {2021})}\BibitemShut {NoStop}%
\bibitem [{\citenamefont {Manzano}\ \emph {et~al.}(2020)\citenamefont {Manzano}, \citenamefont {S\'anchez}, \citenamefont {Silva}, \citenamefont {Haack}, \citenamefont {Brask}, \citenamefont {Brunner},\ and\ \citenamefont {Potts}}]{Ralph}%
  \BibitemOpen
  \bibfield  {author} {\bibinfo {author} {\bibfnamefont {G.}~\bibnamefont {Manzano}}, \bibinfo {author} {\bibfnamefont {R.}~\bibnamefont {S\'anchez}}, \bibinfo {author} {\bibfnamefont {R.}~\bibnamefont {Silva}}, \bibinfo {author} {\bibfnamefont {G.}~\bibnamefont {Haack}}, \bibinfo {author} {\bibfnamefont {J.~B.}\ \bibnamefont {Brask}}, \bibinfo {author} {\bibfnamefont {N.}~\bibnamefont {Brunner}}, \ and\ \bibinfo {author} {\bibfnamefont {P.~P.}\ \bibnamefont {Potts}},\ }\href {\doibase 10.1103/PhysRevResearch.2.043302} {\bibfield  {journal} {\bibinfo  {journal} {Phys. Rev. Res.}\ }\textbf {\bibinfo {volume} {2}},\ \bibinfo {pages} {043302} (\bibinfo {year} {2020})}\BibitemShut {NoStop}%
\bibitem [{\citenamefont {Hwang}\ \emph {et~al.}(2023)\citenamefont {Hwang}, \citenamefont {Sothmann},\ and\ \citenamefont {S\'anchez}}]{Hwarn}%
  \BibitemOpen
  \bibfield  {author} {\bibinfo {author} {\bibfnamefont {S.-Y.}\ \bibnamefont {Hwang}}, \bibinfo {author} {\bibfnamefont {B.}~\bibnamefont {Sothmann}}, \ and\ \bibinfo {author} {\bibfnamefont {D.}~\bibnamefont {S\'anchez}},\ }\href {\doibase 10.1103/PhysRevB.107.245412} {\bibfield  {journal} {\bibinfo  {journal} {Phys. Rev. B}\ }\textbf {\bibinfo {volume} {107}},\ \bibinfo {pages} {245412} (\bibinfo {year} {2023})}\BibitemShut {NoStop}%
\bibitem [{\citenamefont {Tabatabaei}\ \emph {et~al.}(2022)\citenamefont {Tabatabaei}, \citenamefont {S\'anchez}, \citenamefont {Yeyati},\ and\ \citenamefont {S\'anchez}}]{Mojtaba}%
  \BibitemOpen
  \bibfield  {author} {\bibinfo {author} {\bibfnamefont {S.~M.}\ \bibnamefont {Tabatabaei}}, \bibinfo {author} {\bibfnamefont {D.}~\bibnamefont {S\'anchez}}, \bibinfo {author} {\bibfnamefont {A.~L.}\ \bibnamefont {Yeyati}}, \ and\ \bibinfo {author} {\bibfnamefont {R.}~\bibnamefont {S\'anchez}},\ }\href {\doibase 10.1103/PhysRevB.106.115419} {\bibfield  {journal} {\bibinfo  {journal} {Phys. Rev. B}\ }\textbf {\bibinfo {volume} {106}},\ \bibinfo {pages} {115419} (\bibinfo {year} {2022})}\BibitemShut {NoStop}%
\bibitem [{\citenamefont {Beenakker}\ and\ \citenamefont {van Houten}(1991)}]{Beenakker}%
  \BibitemOpen
  \bibfield  {author} {\bibinfo {author} {\bibfnamefont {C.}~\bibnamefont {Beenakker}}\ and\ \bibinfo {author} {\bibfnamefont {H.}~\bibnamefont {van Houten}},\ }\enquote {\bibinfo {title} {Quantum transport in semiconductor nanostructures},}\ in\ \href {\doibase 10.1016/s0081-1947(08)60091-0} {\emph {\bibinfo {booktitle} {Semiconductor Heterostructures and Nanostructures}}}\ (\bibinfo  {publisher} {Elsevier},\ \bibinfo {year} {1991})\ p.\ \bibinfo {pages} {1–228}\BibitemShut {NoStop}%
\bibitem [{\citenamefont {Datta}(1997)}]{Datta}%
  \BibitemOpen
  \bibfield  {author} {\bibinfo {author} {\bibfnamefont {S.}~\bibnamefont {Datta}},\ }\href@noop {} {\emph {\bibinfo {title} {Electronic Transport in Mesoscopic Systems}}}\ (\bibinfo  {publisher} {Cambridge University Press},\ \bibinfo {address} {Cambridge, UK},\ \bibinfo {year} {1997})\BibitemShut {NoStop}%
\bibitem [{\citenamefont {Aharonov}\ and\ \citenamefont {Bohm}(1959)}]{AB}%
  \BibitemOpen
  \bibfield  {author} {\bibinfo {author} {\bibfnamefont {Y.}~\bibnamefont {Aharonov}}\ and\ \bibinfo {author} {\bibfnamefont {D.}~\bibnamefont {Bohm}},\ }\href {\doibase 10.1103/PhysRev.115.485} {\bibfield  {journal} {\bibinfo  {journal} {Phys. Rev.}\ }\textbf {\bibinfo {volume} {115}},\ \bibinfo {pages} {485} (\bibinfo {year} {1959})}\BibitemShut {NoStop}%
\bibitem [{\citenamefont {Haack}\ and\ \citenamefont {Giazotto}(2019)}]{Gérald}%
  \BibitemOpen
  \bibfield  {author} {\bibinfo {author} {\bibfnamefont {G.}~\bibnamefont {Haack}}\ and\ \bibinfo {author} {\bibfnamefont {F.}~\bibnamefont {Giazotto}},\ }\href {\doibase 10.1103/PhysRevB.100.235442} {\bibfield  {journal} {\bibinfo  {journal} {Phys. Rev. B}\ }\textbf {\bibinfo {volume} {100}},\ \bibinfo {pages} {235442} (\bibinfo {year} {2019})}\BibitemShut {NoStop}%
\bibitem [{\citenamefont {Behera}\ \emph {et~al.}(2023)\citenamefont {Behera}, \citenamefont {Bedkihal}, \citenamefont {Agarwalla},\ and\ \citenamefont {Bandyopadhyay}}]{Behera}%
  \BibitemOpen
  \bibfield  {author} {\bibinfo {author} {\bibfnamefont {J.}~\bibnamefont {Behera}}, \bibinfo {author} {\bibfnamefont {S.}~\bibnamefont {Bedkihal}}, \bibinfo {author} {\bibfnamefont {B.~K.}\ \bibnamefont {Agarwalla}}, \ and\ \bibinfo {author} {\bibfnamefont {M.}~\bibnamefont {Bandyopadhyay}},\ }\href {\doibase 10.1103/PhysRevB.108.165419} {\bibfield  {journal} {\bibinfo  {journal} {Phys. Rev. B}\ }\textbf {\bibinfo {volume} {108}},\ \bibinfo {pages} {165419} (\bibinfo {year} {2023})}\BibitemShut {NoStop}%
\bibitem [{\citenamefont {Bedkihal}\ \emph {et~al.}(2025)\citenamefont {Bedkihal}, \citenamefont {Behera},\ and\ \citenamefont {Bandyopadhyay}}]{Bedkiha}%
  \BibitemOpen
  \bibfield  {author} {\bibinfo {author} {\bibfnamefont {S.}~\bibnamefont {Bedkihal}}, \bibinfo {author} {\bibfnamefont {J.}~\bibnamefont {Behera}}, \ and\ \bibinfo {author} {\bibfnamefont {M.}~\bibnamefont {Bandyopadhyay}},\ }\href {\doibase 10.1088/1361-648x/adb921} {\bibfield  {journal} {\bibinfo  {journal} {Journal of Physics: Condensed Matter}\ }\textbf {\bibinfo {volume} {37}},\ \bibinfo {pages} {163001} (\bibinfo {year} {2025})}\BibitemShut {NoStop}%
\bibitem [{\citenamefont {Rashba}(1960)}]{Rashba}%
  \BibitemOpen
  \bibfield  {author} {\bibinfo {author} {\bibfnamefont {E.}~\bibnamefont {Rashba}},\ }\href {https://journals.aps.org/prb/abstract/10.1103/PhysRevB.69.155335} {\bibfield  {journal} {\bibinfo  {journal} {Sov. Phys. Solid State 2}\ } (\bibinfo {year} {1960})}\BibitemShut {NoStop}%
\bibitem [{\citenamefont {Meijer}\ \emph {et~al.}(2002)\citenamefont {Meijer}, \citenamefont {Morpurgo},\ and\ \citenamefont {Klapwijk}}]{Meijer}%
  \BibitemOpen
  \bibfield  {author} {\bibinfo {author} {\bibfnamefont {F.~E.}\ \bibnamefont {Meijer}}, \bibinfo {author} {\bibfnamefont {A.~F.}\ \bibnamefont {Morpurgo}}, \ and\ \bibinfo {author} {\bibfnamefont {T.~M.}\ \bibnamefont {Klapwijk}},\ }\href {\doibase 10.1103/PhysRevB.66.033107} {\bibfield  {journal} {\bibinfo  {journal} {Phys. Rev. B}\ }\textbf {\bibinfo {volume} {66}},\ \bibinfo {pages} {033107} (\bibinfo {year} {2002})}\BibitemShut {NoStop}%
\bibitem [{\citenamefont {Moln\'ar}\ \emph {et~al.}(2004)\citenamefont {Moln\'ar}, \citenamefont {Peeters},\ and\ \citenamefont {Vasilopoulos}}]{Molnár}%
  \BibitemOpen
  \bibfield  {author} {\bibinfo {author} {\bibfnamefont {B.}~\bibnamefont {Moln\'ar}}, \bibinfo {author} {\bibfnamefont {F.~M.}\ \bibnamefont {Peeters}}, \ and\ \bibinfo {author} {\bibfnamefont {P.}~\bibnamefont {Vasilopoulos}},\ }\href {\doibase 10.1103/PhysRevB.69.155335} {\bibfield  {journal} {\bibinfo  {journal} {Phys. Rev. B}\ }\textbf {\bibinfo {volume} {69}},\ \bibinfo {pages} {155335} (\bibinfo {year} {2004})}\BibitemShut {NoStop}%
\bibitem [{\citenamefont {Aharonov}\ and\ \citenamefont {Casher}(1984)}]{AB-Cash}%
  \BibitemOpen
  \bibfield  {author} {\bibinfo {author} {\bibfnamefont {Y.}~\bibnamefont {Aharonov}}\ and\ \bibinfo {author} {\bibfnamefont {A.}~\bibnamefont {Casher}},\ }\href {\doibase 10.1103/PhysRevLett.53.319} {\bibfield  {journal} {\bibinfo  {journal} {Phys. Rev. Lett.}\ }\textbf {\bibinfo {volume} {53}},\ \bibinfo {pages} {319} (\bibinfo {year} {1984})}\BibitemShut {NoStop}%
\bibitem [{\citenamefont {Moskalets}(2011)}]{Moskalets2011}%
  \BibitemOpen
  \bibfield  {author} {\bibinfo {author} {\bibfnamefont {M.~V.}\ \bibnamefont {Moskalets}},\ }\href@noop {} {\emph {\bibinfo {title} {Scattering Matrix Approach to Non-Stationary Quantum Transport}}}\ (\bibinfo {year} {2011})\ Chap.~\bibinfo {chapter} {1},\ \bibinfo {note} {lecture notes}\BibitemShut {NoStop}%
\bibitem [{\citenamefont {Singh}\ and\ \citenamefont {Sahoo}(2025)}]{singh20}%
  \BibitemOpen
  \bibfield  {author} {\bibinfo {author} {\bibfnamefont {K.}~\bibnamefont {Singh}}\ and\ \bibinfo {author} {\bibfnamefont {R.}~\bibnamefont {Sahoo}},\ }\href {https://arxiv.org/abs/2508.00407} {\enquote {\bibinfo {title} {Thermoelectric figure of merit and the deconfinement phase transition},}\ } (\bibinfo {year} {2025}),\ \Eprint {http://arxiv.org/abs/2508.00407} {arXiv:2508.00407 [hep-ph]} \BibitemShut {NoStop}%
\bibitem [{\citenamefont {Okawa}\ \emph {et~al.}(2024)\citenamefont {Okawa}, \citenamefont {Amagai}, \citenamefont {Sakamoto},\ and\ \citenamefont {Kaneko}}]{Okawa_20}%
  \BibitemOpen
  \bibfield  {author} {\bibinfo {author} {\bibfnamefont {K.}~\bibnamefont {Okawa}}, \bibinfo {author} {\bibfnamefont {Y.}~\bibnamefont {Amagai}}, \bibinfo {author} {\bibfnamefont {N.}~\bibnamefont {Sakamoto}}, \ and\ \bibinfo {author} {\bibfnamefont {N.-H.}\ \bibnamefont {Kaneko}},\ }\href {\doibase 10.1016/j.measurement.2024.114626} {\bibfield  {journal} {\bibinfo  {journal} {Measurement}\ }\textbf {\bibinfo {volume} {232}},\ \bibinfo {pages} {114626} (\bibinfo {year} {2024})}\BibitemShut {NoStop}%
\bibitem [{\citenamefont {Khatri}\ \emph {et~al.}(2019)\citenamefont {Khatri}, \citenamefont {Matyas}, \citenamefont {Siddiqui},\ and\ \citenamefont {Dowling}}]{Khat}%
  \BibitemOpen
  \bibfield  {author} {\bibinfo {author} {\bibfnamefont {S.}~\bibnamefont {Khatri}}, \bibinfo {author} {\bibfnamefont {C.~T.}\ \bibnamefont {Matyas}}, \bibinfo {author} {\bibfnamefont {A.~U.}\ \bibnamefont {Siddiqui}}, \ and\ \bibinfo {author} {\bibfnamefont {J.~P.}\ \bibnamefont {Dowling}},\ }\href {\doibase 10.1103/PhysRevResearch.1.023032} {\bibfield  {journal} {\bibinfo  {journal} {Phys. Rev. Res.}\ }\textbf {\bibinfo {volume} {1}},\ \bibinfo {pages} {023032} (\bibinfo {year} {2019})}\BibitemShut {NoStop}%
\bibitem [{\citenamefont {Kim}\ \emph {et~al.}(2015)\citenamefont {Kim}, \citenamefont {Liu}, \citenamefont {Chen}, \citenamefont {Chu},\ and\ \citenamefont {Ren}}]{Kim}%
  \BibitemOpen
  \bibfield  {author} {\bibinfo {author} {\bibfnamefont {H.~S.}\ \bibnamefont {Kim}}, \bibinfo {author} {\bibfnamefont {W.}~\bibnamefont {Liu}}, \bibinfo {author} {\bibfnamefont {G.}~\bibnamefont {Chen}}, \bibinfo {author} {\bibfnamefont {C.-W.}\ \bibnamefont {Chu}}, \ and\ \bibinfo {author} {\bibfnamefont {Z.}~\bibnamefont {Ren}},\ }\href {\doibase 10.1073/pnas.1510231112} {\bibfield  {journal} {\bibinfo  {journal} {Proceedings of the National Academy of Sciences}\ }\textbf {\bibinfo {volume} {112}},\ \bibinfo {pages} {8205} (\bibinfo {year} {2015})},\ \Eprint {http://arxiv.org/abs/https://www.pnas.org/doi/pdf/10.1073/pnas.1510231112} {https://www.pnas.org/doi/pdf/10.1073/pnas.1510231112} \BibitemShut {NoStop}%
\bibitem [{\citenamefont {Beretta}(2024)}]{Bereta}%
  \BibitemOpen
  \bibfield  {author} {\bibinfo {author} {\bibfnamefont {D.}~\bibnamefont {Beretta}},\ }\href {https://arxiv.org/abs/2412.14885} {\enquote {\bibinfo {title} {On the strategies to enhance zt},}\ } (\bibinfo {year} {2024}),\ \Eprint {http://arxiv.org/abs/2412.14885} {arXiv:2412.14885 [cond-mat.mtrl-sci]} \BibitemShut {NoStop}%
\bibitem [{\citenamefont {Whitney}\ \emph {et~al.}(2018)\citenamefont {Whitney}, \citenamefont {S{\'a}nchez},\ and\ \citenamefont {Splettstoesser}}]{Rafael}%
  \BibitemOpen
  \bibfield  {author} {\bibinfo {author} {\bibfnamefont {R.~S.}\ \bibnamefont {Whitney}}, \bibinfo {author} {\bibfnamefont {R.}~\bibnamefont {S{\'a}nchez}}, \ and\ \bibinfo {author} {\bibfnamefont {J.}~\bibnamefont {Splettstoesser}},\ }\enquote {\bibinfo {title} {Quantum thermodynamics of nanoscale thermoelectrics and electronic devices},}\ in\ \href {\doibase 10.1007/978-3-319-99046-0_7} {\emph {\bibinfo {booktitle} {Thermodynamics in the Quantum Regime: Fundamental Aspects and New Directions}}},\ \bibinfo {editor} {edited by\ \bibinfo {editor} {\bibfnamefont {F.}~\bibnamefont {Binder}}, \bibinfo {editor} {\bibfnamefont {L.~A.}\ \bibnamefont {Correa}}, \bibinfo {editor} {\bibfnamefont {C.}~\bibnamefont {Gogolin}}, \bibinfo {editor} {\bibfnamefont {J.}~\bibnamefont {Anders}}, \ and\ \bibinfo {editor} {\bibfnamefont {G.}~\bibnamefont {Adesso}}}\ (\bibinfo  {publisher} {Springer International Publishing},\ \bibinfo {address} {Cham},\ \bibinfo {year} {2018})\ pp.\ \bibinfo {pages} {175--206}\BibitemShut
  {NoStop}%
\end{thebibliography}%
